\documentclass[12pt]{iopart}

\usepackage{graphicx}

\usepackage{iopams}
\usepackage{amsfonts}

\begin{document}

\title[Dirac particle]
{Classical Dirac particle I}

\author{Juan Barandiaran}
\address{Telecommunication Engineer, Bilbao, Spain}
\ead{barandiaran.juan@gmail.com}
\author{Mart\'{\i}n Rivas}
\address{Theoretical Physics Department, The University of the Basque Country, (retired)\\ 
Bilbao, Spain}
\ead{martin.rivas@ehu.eus}

\begin{abstract}
In this work we produce a classical Lagrangian description of an elementary spinning particle which satisfies Dirac equation when quantized. We call this particle a {\bf classical Dirac particle}. We analyze in detail the way we arrive to this model and how the different observables and constants of the motion can be expressed in terms of the degrees of freedom and their derivatives, by making use of Noether's theorem. The main feature is that the particle has a center of charge ${\bi r}$, moving at the speed of light, that satisfies fourth-order differential equations and all observables can be expressed only in terms of this point and their time derivatives. The particle has also a center of mass ${\bi q}$, that is a different point than the center of charge. This implies that two different spin observables can be defined, one ${\bi S}$ with respect to the point ${\bi r}$ and another ${\bi S}_{CM}$ with respect to the point ${\bi q}$, that satisfy different dynamical equations. The spin ${\bi S}$ satisfies the same dynamical equation than Dirac's spin operator. The fourth-order differential equations for the point ${\bi r}$ can be transformed into a system of second-order ordinary differential equations for the center of charge and center of mass. The dynamics can be described in terms of dimensionless variables in the natural system of units. We also obtain the dynamical equations of the center of mass spin  ${\bi S}_{CM}$. The possible interaction Lagrangians are described and we devote the main part of the work to the electromagnetic interaction of the Dirac particle with uniform and oscillating electric and magnetic fields. The numerical integrations of the dynamical equations are performed with different Mathematica notebooks that are available for the interested reader.

\end{abstract}

\hspace{1.2cm}{\small\bf Keywords:} {Spinning electron; Dirac particle}

\section{Introduction}
\label{intro}

We call a classical Dirac particle a classical mechanical system obtained through a general formalism to describe classical elementary particles with spin, that satisfies the Dirac equation when it is quantized. 

It depends on two intrinsic parameters: the mass $m$ and the absolute value $S$ of the spin in the center of mass frame. When quantized, the value of the nonvanishing mass $m$ is arbitrary but the value of $S=\hbar/2$, and it represents a classical model of a fermion.

The main feature of this classical spinning model is that we only need to describe the evolution of a single point ${\bi r}$, the center of charge, where the external forces are defined, and which satisfies a system of fourth-order differential equations. All observables, energy, linear momentum, angular momentum, center of mass position, angular velocity and spins are expressed in terms of this point ${\bi r}$ and their time derivatives. 

This model has also another characteristic point, the center of mass ${\bi q}$, which is a different point than the center of charge. The fourth-order differential equations for the point ${\bi r}$ can be decoupled into a system of second-order differential equations for the points ${\bi q}$ and ${\bi r}$, where ${\bi q}$ satisfies Newton-type dynamical equations in terms of the external force evaluated at the position of the center of charge ${\bi r}$.

The center of charge ${\bi r}$ is moving at the speed of light and the complete dynamical equations for the evolution of the center of mass ${\bi q}$ and center of charge ${\bi r}$ can be expressed in terms of dimensionless classical variables. 

In the Section {\bf\ref{summary}} we make a summary of the theoretical ideas that have lead us to this formalism and which are developped in the mentioned Sections and references.

\section{Scope and steps of the theoretical analysis}
\label{summary}

The formalism we have proposed is based on the following considerations \cite{Rivasbook}:
\begin{enumerate}
\item{The constituent elements of matter are elementary particles.}
\item{From the point of view of classical mechanics it is possible to describe an elementary particle as a mechanical Lagrangian system with a finite but unknown number of degrees of freedom. Section {\bf\ref{clasicalelectron}}.}
\item{In nature, there is a limiting speed $c$ for the motion of the centers of mass of massive bodies.
This implies that equivalent inertial observers are related by the Poincar\'e group transformations. Section {\bf\ref{clasicalelectron}}.}
\item{We accept that elementary particles interact. We assume that when an elementary particle interacts and if it is not annihilated, its internal structure is not modified by the interaction. Therefore, the boundary variables of the Lagrangian description of an elementary particle are restricted to belong to a homogenenous space of the Poincar\'e group \cite{atomic}. Section {\bf\ref{clasicalelectron}}. The maximum number of boundary variables is the same as the parameters of the Poincar\'e group.}
\item{The formalism must be quantized and our interest is to find at least one model that satisfies the Dirac equation when quantized.  \cite{RivasDirac}.}
\item{The Lagrangian that describes the more complex elementary particle has as boundary variables $(t,{\bi r},{\bi u},\balpha)$, what are interpreted as the time $t$, the position of a point ${\bi r}$, the velocity of this point ${\bi u}$ and the orientation $\balpha$ of a comoving cartesian frame attached to the point ${\bi r}$. It is therefore a function $L((t,{\bi r},{\bi u},{\bi a},\bomega)$ of these variables and of the acceleration of the point ${\bi a}$ and of the angular velocity $\bomega$ of the comoving frame. The free Lagrangian $L_0({\bi u},{\bi a},\bomega)$, is invariant under translations and, therefore, independent of $t$ and ${\bi r}$. Section {\bf \ref{Homogeneity}}.}
\item{The point ${\bi r}$ must satisfy fourth-order differential equations, so its free motion is accelerated and has torsion and therefore does not represent the center of inertia or center of mass (CM) of the free particle. The possible interaction Lagrangians will be defined at that point. The point ${\bi r}$ represents the center of interaction or center of charge (CC) of the elementary particle. Section {\bf \ref{Homogeneity}}.}
\item{The classical model we have found that when quantized satisfies the Dirac equation corresponds to the previous system in which $u=c$, i.e., the CC moves at the speed of light. We call this elementary particle the classical Dirac particle. \cite{Rivasbook}, \cite{RivasDirac}.}
\item{Without explicit knowledge of the free Lagrangian and by using Noether's theorem we have found that the free particle has another characteristic point ${\bi q}$, defined in terms of ${\bi r}$ and its time derivatives and that for the free particle moves in a straight line at a constant speed ${\bi v}=d{\bi q}/dt$. The energy $H$ and the linear momentum ${\bi p}$ of the particle are expressed in terms of this velocity in the form $H=\gamma(v)mc^2$ and ${\bi p}=\gamma(v)m{\bi v}$. We interpret this point as the center of inertia or center of mass of the particle. We have found the differential equations that satisfy the point ${\bi r}$ for the center of mass observer who measures  ${\bi q}=0$ and ${\bi v}=0$. Section {\bf\ref{constants}}.}
\item{Because the Dirac particle has two distinguished points ${\bi q}$ and  ${\bi r}$ we obtain two different angular momenta observables with respect to them. They are the spin ${\bi S}$ with respect to the CC and the spin ${\bi S}_{CM}$ with respect to the CM. Both spins satisfy different dynamical equations. All observables of the Dirac particle can finally be expressed in terms of the point ${\bi r}$ and its subsequent time derivatives. Section {\bf\ref{spins}}.}
\item{If the trajectory of the point ${\bi r}$  for the center of mass observer is written in another inertial reference system by means of an arbitrary Poincar\'e transformation, we obtain the complete family of all possible trajectories of the point described by all inertial observers. If in that family of trajectories we eliminate the parameters of the Poincar\'e group on which it depends, from the equation of the family and its successive time derivatives we finally obtain a fourth-order system of differential equations for the point ${\bi r}$, independent of the parameters of the group. These are the Poincar\'e invariant relativistic differential equations of the free particle. \cite{RivasDynamics} }
\item{By the analysis of the Poincar\'e invariant magnitudes of the Dirac particle we have found how the spin transforms among inertial observers and the dynamical equations of the center of mass spin. Section {\bf\ref{PoinInvariant}}.}\item{From the expression of ${\bi q}$ in terms of ${\bi r}$ and its time derivatives it has been possible to transform the system of fourth-order differential equations for the point ${\bi r}$, in a coupled second-order system of ordinary differential equations for the points ${\bi q}$ and ${\bi r}$, and extend these equations when there exist any interaction. Section {\bf\ref{spins}}.}
\item{The most general interaction Lagrangian of the Dirac particle is described by  $L=L_0+L_I$, where
$L_0({\bi u},{\bi a},\bomega)$ is the still unknown free Lagrangian and the interaction Lagrangian $L_I(t,{\bi r},{\bi u})=A_0(t,{\bi r},{\bi u})+{\bi u}\cdot{\bi A}(t,{\bi r},{\bi u})$, is only a function of $t,{\bi r}$ and ${\bi u}$, independent of ${\bi a}$ and $\bomega$, as a consequence that the interaction does not modify the internal structure of the Dirac particle. Section {\bf\ref{InterLag}}.}
\item{The possibility that the functions $A_0$ and  ${\bi A} $ may also be functions of the CC speed ${\bi u}$ has not been considered here, and opens the door to the analysis of possible classical interactions of non-electromagnetic nature that should be explored. Section {\bf\ref{interaction}}.} 
\item{We have expressed the dynamical equations in natural units where the unit of velocity is $c$, the unit of length is $2R_0$, being $R_0$ the separation between the CC and CM in the center of mass frame and the unit of time is $2R_0/c$. The boundary conditions for solving the dynamical equations are expressed in terms of 10 essential parameters. These essential parameters define the relationship between the laboratory observer and the center of mass observer of the Dirac particle. Section {\bf\ref{Natural}}.}
\item{In this work we have analyzed the case in which the interaction is of the form of the {\it minimal coupling},
$L_I=-eA_0(t,{\bi r})+e{\bi u}\cdot{\bi A}(t,{\bi r})$, in terms of the electric charge and in terms of the scalar  $A_0$, and vector potential ${\bi A}$, defined at the CC position ${\bi r}$, which give rise to uniform and oscillating electric and magnetic fields of constant frequency. The surprise is that we do not need the explicit knowledge of the free Lagrangian. Sections {\bf\ref{sec:uniformE}-\ref{sec:uniformB}}.}

\end{enumerate}

\section{Classical elementary spinning particles}
\label{clasicalelectron}

Different models of classical elementary spinning particles have been obtained through a general formalism \cite{Rivasbook}, that is based on the following three fundamental principles: Restricted Relativity Principle, Variational Principle and Atomic Principle. 

The {\bf Restricted Relativity Principle} admits the existence of a class of equivalent observers which describe the mechanical systems with the same dynamical equations. Equivalent observers are related by means of a group of space-time transformations, called the kinematical group of the formalism. For describing the Dirac particle we shall restrict this group to the Poincar\'e group. We call this principle as a {\bf restricted} principle because we are going to consider only the class of inertial observers. Accelerated observers are not considered here.

The {\bf Variational Principle} states that the dynamical equations of any mechanical system between an initial fixed point $x_1$ to a final fixed point $x_2$ of a kinematical space $x_1,x_2\in X$, are  the Euler-Lagrange dynamical equations obtained from a Lagrangian function which is an explicit function of the variables $x$ and their first order time derivative $L(x,\dot{x})$. 

The definition of a classical elementary particle lies on the {\bf Atomic principle} \cite{atomic}. The assumption of this principle is that an elementary particle does not have internal excited states in the sense that any interaction, if it does not annihilate the particle, does not modify its internal structure. If an inertial observer describes the initial state of the elementary particle in the Lagrangian description through a set of variables $x\equiv(x_1,\ldots,x_n)$ and the dynamics changes this state to  another $y\equiv(y_1,\ldots,y_n)$, if the internal structure has not been modified, then it is possible to find another inertial observer who describes this new state of the particle with exactly the same values of the variables as of the previous state. This means that there must exist a Poincar\'e transformation $g$, between both inertial observers, in such a way that the above values are transformed into each other: 
\[
y=g*x,\quad x=g^{-1}*y.
\]
This relation must be valid for any pair of states of any elementary particle.
The Atomic Principle restricts the classical boundary variables manifold of the variational description of an elementary particle, to span a homogeneous space of the Poincar\'e group \cite{Rivasbook}. 
The largest homogeneous space of a Lie group is the group itself, and the possible different homogeneous spaces of the Poincar\'e group will supply different models of elementary particles. 
The point particle is the simplest of these possible models.

\subsection{The boundary variables}
If $t$ and ${\bi r}$ are the time and position coordinates of a space-time event for the inertial observer $O$, and $t'$ and ${\bi r}'$ are the time and position of the same event for any arbitrary inertial observer $O'$, they are related by the Poincar\'e transformation:
\[
{x^{\mu}}'=\Lambda^\mu_\nu x^{\nu}+a^\mu,\quad {x^{\mu}}'\equiv(ct',{\bi r}'),\quad x^{\mu}\equiv(ct,{\bi r}), \quad a^\mu\equiv(cb,{\bi d}),
\]
and $\Lambda=L({\bi v})R(\balpha)$ is a general Lorentz transformation, as a composition of a rotation $R(\balpha)$, followed by a boost or pure Lorentz transformation $L({\bi v})$. The three vector $\balpha$ that defines the rotation matrix can be the vector
$\balpha=\alpha{\bi n}$, where ${\bi n}$ is a unit vector along the rotation axis and $\alpha$ the rotated angle around this direction, so that any rotation is finally described in terms of three angles, $\alpha$ and the zenithal $\theta$ and azimuthal $\phi$ that describe the unit vector ${\bi n}$.

A Lorentz transformation of velocity ${\bi v}$, is given by the $4\times4$ matrix
 \begin{equation}
L({\bi v})=\pmatrix{\gamma&\gamma{v_x/ c}& \gamma{v_y/ c}& 
\gamma{v_z/c}\cr 
\gamma{v_x/c}&1+{\displaystyle v_x^2\over 
\displaystyle c^2}{\displaystyle\gamma^2\over\displaystyle\gamma+1}&{\displaystyle v_xv_y\over\displaystyle 
c^2}{\displaystyle\gamma^2\over\displaystyle\gamma+1}&{\displaystyle v_xv_z\over\displaystyle c^2}{\displaystyle\gamma^2\over\displaystyle\gamma+1}\cr 
\gamma{v_y/ c}&{\displaystyle v_yv_x\over\displaystyle c^2}{\displaystyle\gamma^2\over\displaystyle\gamma+1}& 1+{\displaystyle v_y^2\over\displaystyle 
\displaystyle c^2}{\displaystyle\gamma^2\over\displaystyle 
\gamma+1}&{\displaystyle v_yv_z\over\displaystyle 
c^2}{\displaystyle\gamma^2\over\displaystyle\gamma+1}\cr 
\gamma{v_z/ c}&{\displaystyle v_zv_x\over 
\displaystyle c^2}{\displaystyle\gamma^2\over\displaystyle \gamma+1}&{\displaystyle v_zv_y\over 
\displaystyle c^2}{\displaystyle\gamma^2\over\displaystyle \gamma+1}&1+{\displaystyle v_z^2\over 
\displaystyle c^2}{\displaystyle\gamma^2\over\displaystyle \gamma+1}\cr},
 \label{eq:Tdev}
 \end{equation} 
where $\gamma\equiv(1-(v_x/c)^2-(v_y/c)^2-(v_z/c)^2)^{-1/2}$. The Poincar\'e transformation is
 \begin{eqnarray}
t'&=&\gamma t+\gamma({\bi v}\cdot R(\balpha){\bi r})/c^2+b,\qquad\label{eq:5.1a}\\ 
{\bi r}'&=&R(\balpha){\bi r}+\gamma{\bi v}t+\frac{\gamma^2}{(1+\gamma)c^2}
({\bi v}\cdot R(\balpha){\bi r})){\bi v}+{\bi d},\qquad\label{eq:5.1b}   
 \end{eqnarray} 

The event $(0,{\bf 0})$ in $O$ is described by $(b,{\bi d})$ in $O'$, so that these values represent the time and position of the origin of $O$, measured by $O'$, when the clock of $O$ marks zero. The velocity ${\bi v}$ is the velocity of the origin of $O$ as measured by $O'$. Finally $R(\balpha)$ is the rotation $O'$ has to perform to its own Cartesian frame to obtain the Cartesian frame of $O$ before the Lorentz boost $L({\bi v})$. The variables $(b,{\bi d},{\bi v},\balpha)$ of a Poincar\'e group element is the way the observer $O'$ describes any other observer $O$. This is the maximal set of variables we can use to describe the boundary variables of any elementary particle. We do not make any assumption about the size or shape of the particle.

Then the initial state of an elementary particle is characterized at most by the ten variables $x_i\equiv(t,{\bi r},{\bi u},\balpha)$, that are interpreted as the time $t$, the position of a single point ${\bi r}$, the velocity of this point ${\bi u}$,
and finally the orientation $\balpha\in SO(3)$, of a comoving Cartesian frame attached to the point ${\bi r}$. The same variables, with different values, for the final point $x_f$ of the variational evolution. 
If we restrict ourselves to a system where the initial state is given by the variables $x_i\equiv(t,{\bi r})$, which span a homogeneous space of the Poincar\'e group, we are describing another model of elementary particle: the spinless point particle. It is in terms of the extra variables $({\bi u},\balpha)$, that we will be able to describe the spin structure. We shall call {\bf kinematical variables} to the boundary variables $(t,{\bi r},{\bi u},\balpha)$.
Elementary particles are localized by a single point, and they move and rotate.

What we see is that we have three different manifolds spanned by the variables $(t,{\bi r},{\bi u},\balpha)$, provided the variable $u<c$, $u=c$ or $u>c$.
Among the models of spinning particles this formalism predicts, the only one that satisfies Dirac's equation when quantized \cite{RivasDirac}, is that model in which the point ${\bi r}$, is moving at the speed of light. 

Since the Lagrangian also depends  on the next order time derivative of the above boundary variables, the Lagrangian must depend on the acceleration of the point ${\bi a}$, and on the angular velocity $\bomega$. The point ${\bi r}$, satisfies therefore  a system of fourth-order differential equations and, being the only point where the external fields will be defined, we interpret the point ${\bi r}$, as the location of the center of charge (CC) of the particle.

\section{Homogeneity of the Lagrangian}
\label{Homogeneity}
The most general Lagrangian for describing a mechanical system with boundary variables
$x$  will be an explicit function of these variables and their next order time derivative $L(x,dx/dt)$.
The usual way to state the variation of the action is to define the action functional
\[
A[x(t)]=\int_{t_1}^{t_2}L(x(t),dx(t)/dt)\,dt,
\]
for any arbitrary path $x(t)$ in $X$ space between the fixed initial and final states $x(t_1)$ and $x(t_2)$, respectively.
But time is a relative observable to some particular observer. To produce a formalism independent of the observer let us assume that the evolution can be described in terms of some arbitrary evolution parameter $\tau$, the same for all inertial observers. Then we assume that all variables are continuous and differentiable functions of this parameter $x(\tau)$.
The action functional is now written as
\[
A[x(\tau)]=\int_{t_1}^{t_2}L(x,dx/dt)\,dt=\int_{\tau_1}^{\tau_2}L(x(\tau),\dot{x}/\dot{t})\,\dot{t}(\tau)d\tau
=\int_{\tau_1}^{\tau_2}\widetilde{L}(x,\dot{x})d\tau,
\]
and inside the function $L$, the derivatives $dx/dt$ are replaced by $\dot{x}/\dot{t}$, where the dot represents the derivative with respect to the evolution parameter $\tau$. The function $L$ depends on the $x$ variables and of the quotients $\dot{x}/\dot{t}$, so that it is a homogeneous function of zero-th degree of all the $\dot{x}$. The function $\widetilde{L}=L\dot{t}$ is thus a homogeneous function of first degree in terms of all $\dot{x}$, including $\dot{t}$. This homogeneity allows us to write the Lagrangian as a sum of as many terms as kinematical variables. Euler's theorem
on homogeneous functions $F$, of n-th degree gives
\[
F(x_1,\ldots,x_k)\quad\Rightarrow\quad \sum_{i=1}^{i=k}\frac{\partial F}{\partial x_i}x_i=nF(x_1,\ldots,x_k).
\]

The most general Lagrangian with boundary variables $x\equiv(t,{\bi r},{\bi u},\balpha)\in X$, will be an explicit function of these variables and their next order time derivative. The Lagrangian will be a function $L(x,dx/dt)\equiv L(t,{\bi r},{\bi u},{\bi a},\bomega)$, where the velocity ${\bi u}=d{\bi r}/dt$, the acceleration ${\bi a}=d{\bi u}/dt$ and the dependence on $\balpha$ and $d\balpha/dt$ is through the definition of the angular velocity $\bomega$ which is a function of $\balpha$ and linear in $d\balpha/dt$. 

For the function $\widetilde{L}$, homogeneous of the $\dot{x}$ of degree $n=1$,  we get
\begin{equation}
\widetilde{L}(x,\dot{x})=\sum_{i=1}^{i=k}\frac{\partial \widetilde{L}}{\partial \dot{x}_i}\dot{x}_i.
\label{homogeneity}
\end{equation}
Explicitely we shall write the most general Lagrangian in the form:
\[
\widetilde{L}=T\dot{t}+{\bi R}\cdot\dot{\bi r}+{\bi U}\cdot\dot{\bi u}+{\bi V}\cdot\dot{\balpha},
\]
where $T=\partial\widetilde{L}/\partial\dot{t}$, $R_i=\partial\widetilde{L}/\partial\dot{r}_i$,
$U_i=\partial\widetilde{L}/\partial\dot{u}_i$, and $V_i=\partial\widetilde{L}/\partial\dot{\alpha}_i$. The last term, because $\bomega$ is a linear function of $\dot{\balpha}$, can be replaced by
${\bi W}\cdot\bomega$, where $ W_i=\partial\widetilde{L}/\partial\omega_i$. When selecting some particular observer, the Lagrangian is written in a time evolution description and will be $L=\widetilde{L}/\dot{t}$, or to consider that $\tau=t$, so that
\[
L=T+{\bi R}\cdot{\bi u}+{\bi U}\cdot{\bi a}+{\bi W}\cdot{\bomega},
\]
and also $R_i=\partial{L}/\partial {u}_i$,
$U_i=\partial{L}/\partial{a}_i$, and $W_i=\partial{L}/\partial{\omega_i}$.

As we shall see, the Noether constants of the motion will be expressed in terms of the functions
$T$, ${\bi R}$, ${\bi U}$ and ${\bi W}$, which are still unknown.\footnote{\footnotesize{We represent 3D vector observables in boldface. Expressions like ${\bi a}=\partial L/\partial {\bi b}$, have to be interpreted throughout this work as $a_i=\partial L/\partial b_i$, $i=1,2,3$.}}

If we describe the evolution of a point in terms of $\tau$, the transformation of the variables $t(\tau)$ and ${\bi r}(\tau)$ between inertial observers $O$ and $O'$, at any $\tau$, is
 \begin{eqnarray}
t'(\tau)&=&\gamma t(\tau) +\gamma({\bi v}\cdot R(\balpha){\bi r}(\tau))/c^2+b,\\ 
{\bi r}'(\tau)&=&R(\balpha){\bi r}(\tau)+\gamma{\bi v}t(\tau)+\frac{\gamma^2}{(1+\gamma)c^2}
({\bi v}\cdot R(\balpha){\bi r}(\tau)){\bi v}+{\bi d}.
\end{eqnarray}
Taking the $\tau-$derivative of all terms we get
\[
\dot{t}'(\tau)=\gamma \dot{t}(\tau) +\gamma({\bi v}\cdot R(\balpha)\dot{\bi r}(\tau))/c^2,
\]
\[
\dot{\bi r}'(\tau)=R(\balpha)\dot{\bi r}(\tau)+\gamma{\bi v}\dot{t}(\tau)+\frac{\gamma^2}{(1+\gamma)c^2}
({\bi v}\cdot R(\balpha)\dot{\bi r}(\tau)){\bi v}.
\]
The quotient $\dot{\bi r}/\dot{t}=d{\bi r}/dt={\bi u}$ and $\dot{\bi r}'/\dot{t}'=d{\bi r}'/dt'={\bi u}'$ is the velocity of the point and it transforms among inertial observers:
\[
{\bi u}'(\tau)=\frac{R(\balpha){\bi u}(\tau)+\gamma{\bi v}+\frac{\gamma^2}{(1+\gamma)c^2}
({\bi v}\cdot R(\balpha){\bi u}(\tau)){\bi v}}{\gamma\left(1 +{\bi v}\cdot R(\balpha){\bi u}(\tau)/c^2\right)}.
\]
From this we get
\[
u'^2(\tau)=\frac{u^2(\tau)-c^2}{\gamma^2\left(1 +{\bi v}\cdot R(\balpha){\bi u}(\tau)/c^2\right)^2}+c^2,
\]
so that if $u<c$, $u'<c$, if $u>c$, $u'>c$, and if $u=c$, $u'=c$. We have shown in  \cite{RivasDirac} that the system with $u=c$ satisfies Dirac equation when quantized. The point ${\bi r}$ is moving at the speed $c$ for all inertial observers. It is this constraint we shall use in this article. Since the acceleration ${\bi a}=d{\bi u}/dt=\dot{\bi u}/\dot{t}$, if we call $R({\bi a})=R(\balpha){\bi a}$ and  $R({\bi u})=R(\balpha){\bi u}$ , it transforms under the Poincar\'e group as:
\begin{equation}
{\bi a}'=\frac{(1 +{\bi v}\cdot R({\bi u})/c^2)R({\bi a})-({\bi v}\cdot R({\bi a})/c^2)R({\bi u})-{\displaystyle\frac{\gamma({\bi v}\cdot R({\bi a})){\bi v}}{(1+\gamma)c^2}}}{\gamma^2\left(1 +{\bi v}\cdot R({\bi u})/c^2\right)^3}.
\label{eq:transaccel}
\end{equation}
\section{Constants of the motion}
\label{constants}

The kinematical variables which define the boundary variables of the variational description of an elementary particle are
$(t,{\bi r},{\bi u},\balpha)$. The generalized coordinates are thus the same variables with the time variable excluded. Therefore there will be a canonical conjugate momentum of each of the variables
${\bi r}$, ${\bi u}$ and $\balpha$. The conjugate momenta will be: The conjugate momentum of ${\bi r}$ variables is the vector ${\bi p}_r=\partial L/\partial{\bi u}-d(\partial L/\partial{\bi a})/dt={\bi R}-d{\bi U}/dt$. The conjugate momentum of ${\bi u}$ variables is the vector ${\bi p}_u=\partial L/\partial{\bi a}={\bi U}$, and finally the conjugate momentum of ${\balpha}$ variables is the vector ${\bi p}_\alpha=\partial L/\partial{\bomega}={\bi W}$. The Hamiltonian of this particle is:
\[
H={\bi p}_r\cdot{\bi u}+{\bi p}_u\cdot{\bi a}+{\bi p}_\alpha\cdot{\bomega}-L=-T-{\bi u}\cdot\frac{d{\bi U}}{dt}.
\]
If under some one-dimensional infinitesimal transformation of infinitesimal parameter $\delta g$, of a symmetry group that leave invariant the dynamical equations, the kinematical variables transform as:
\[
t'=t+M_0(x)\delta g,\quad {\bi r}'={\bi r}+{\bi M}_r(x)\delta g,\quad
{\bi u}'={\bi u}+{\bi M}_u(x)\delta g,\quad{\balpha}'={\balpha}+{\bi M}_\alpha(x)\delta g,
\]
where the $M_i(x)$ are functions of the kinematical variables, the corresponding Noether constant of the motion is:
\[
N=HM_0-{\bi p}_r\cdot{\bi M}_r-{\bi p}_u\cdot{\bi M}_u-{\bi p}_\alpha\cdot{\bi M}_\alpha.
\]
If the particle is free, the Lagrangian $L_0$, will be invariant under all transformations of the Poincar\'e group. If we analyze the constants of the motion under infinitesimal time translation, space translation, Lorentz boost and rotation, respectively, we find the following constants of the motion to which we give the names indicated below \cite{Rivasbook}:
 \begin{eqnarray} 
\hbox{\rm temporal momentum}\qquad H&=&-T-{\bi u}\cdot\frac{d{\bi U}}{ dt},\label{eq:n71}\\ 
\hbox{\rm linear momentum}\quad\quad\quad\;\; {\bi P}&=&{\bi R}-\frac{d{\bi U}}{ dt}={\bi p}_r,\label{eq:n72}\\ 
\hbox{\rm kinematical momentum}\quad {\bi K}&=&H{\bi r}/c^2-{\bi P}t-{\bi S}\times{\bi u}/c^2,\label{eq:n73}\\ 
\hbox{\rm angular momentum}\quad\quad\quad {\bi J}&=&{\bi r}\times{\bi P}+{\bi u}\times{\bi U}+{\bi W}={\bi r}\times{\bi P}+{\bi S}.
\label{eq:n7} 
 \end{eqnarray} 
The temporal momentum is the energy or Hamiltonian $H$, and ${\bi S}={\bi u}\times{\bi U}+{\bi W}$ represents the angular momentum of the particle with respect to the point ${\bi r}$, because ${\bi J}$ is the angular momentum of the particle with respect to the origin of the observer frame. ${\bi S}$ is the spin of the system with respect to the point ${\bi r}$. 

The free invariant Lagrangian $L_0$ is independent of $t$ and ${\bi r}$,
\[
L_0=T_0+{\bi R}_0\cdot{\bi u}+{\bi U}_0\cdot{\bi a}+{\bi W}_0\cdot{\bomega},
\]
and the functions $T_0$, ${\bi R}_0$, ${\bi U}_0$ and ${\bi W}_0$, 
are still unknown functions of $({\bi u},{\bi a},\bomega)$.

\section{The center of mass and the spins of the particle}
\label{spins}
If in the expression of the kinematical momentum (\ref{eq:n73}) we define  a position vector  ${\bi k}$ by
\begin{equation}
\frac{1}{c^2}{\bi S}\times{\bi u}=\frac{1}{c^2}H{\bi k},
\label{eq:k}
\end{equation}
the kinematical momentum can be rewritten as
\begin{equation}
{\bi K}=H({\bi r}-{\bi k})/c^2-{\bi P}t=H{\bi q}/c^2-{\bi P}t, \quad {\bi q}={\bi r}-{\bi k},
\label{eq:K2}
\end{equation}
where ${\bi q}$ represents the position of a different point than ${\bi r}$. Taking the time derivative of the constant ${\bi K}$ we get
\[
{\bi P}=\frac{1}{c^2}H{\bi v},\quad {\bi v}=\frac{d{\bi q}}{dt},
\]
and the linear momentum is written in terms of the velocity of the point ${\bi q}$. Since the invariant and constant of the motion $H^2-c^2{\bi P}^2=m^2c^4$ defines the mass of the particle, using the above expression of ${\bi P}$ we obtain that $H$ and ${\bi P}$ take the following form in terms of the constant velocity ${\bi v}$ of the point ${\bi q}$:
\begin{equation}
H=\gamma(v)mc^2,\quad {\bi P}=\gamma(v)m{\bi v},\quad \gamma(v)=(1-v^2/c^2)^{-1/2}.
\label{defHyp}
\end{equation}

The point ${\bi q}$ represents the location of the center of mass (CM) of the particle and it is a different point than the point ${\bi r}$. The energy and linear momentum take the same expressions like the point particle in terms of the CM velocity. When we shall analyze the interaction of the particle, the interaction Lagrangian will be defined at the point ${\bi r}$, so that we consider that this point represents the location of the center of charge (CC).

If we make the cross product with ${\bi u}$ of the expression (\ref{eq:k}) we get
\[
{\bi u}({\bi u}\cdot{\bi S})-c^2{\bi S}=H({\bi r}-{\bi q})\times{\bi u}.
\]
As we shall see in (\ref{sdudtu}) the velocity ${\bi u}$ and ${\bi S}$ are orthogonal vectors and we obtain the explicit expression of the spin with respect to the CC
\begin{equation}
{\bi S}=-\gamma(v)m({\bi r}-{\bi q})\times{\bi u}.
\label{eq:espinS}
\end{equation}
Because we have another point ${\bi q}$, the spin with respect to the CM is defined by
\begin{equation}
{\bi S}_{CM}={\bi S}+({\bi r}-{\bi q})\times{\bi P}=-\gamma(v)m({\bi r}-{\bi q})\times({\bi u}-{\bi v}).
\label{eq:espinSCM}
\end{equation}
The total angular momentum ${\bi J}$ of the particle with respect to the origin of any arbitrary inertial frame, can be written in two different ways as:
\begin{equation}
{\bi J}={\bi S}+{\bi r}\times{\bi P}={\bi S}_{CM}+{\bi q}\times{\bi P}.
\label{eq:J2}
\end{equation}
Both spins satisfy different dynamical equations. If the particle is free, ${\bi v}$ is constant and $d{\bi J}/dt=0$, and $d{\bi P}/dt=0$, leads to
\begin{equation}
\frac{d{\bi S}}{dt}={\bi P}\times{\bi u},\quad \frac{d{\bi S}_{CM}}{dt}=0.
\label{eq:spinDyn}
\end{equation}
The CM spin ${\bi S}_{CM}$, is conserved while the spin with respect to the CC, ${\bi S}$, satisfies the same dynamical equation as Dirac's spin operator in the quantum case. Its time derivative is always orthogonal to the direction of the linear momentum. We shall see in Section {\bf\ref{angular}} that
${\bi S}$ precesses around the center of mass spin ${\bi S}_{CM}$.

Taking the time derivative of the kinematical momentum (\ref{eq:n73}) we get, because in the free case $H$ and ${\bi P}$ are constants of the motion:
\begin{equation}
\frac{d{\bi K}}{dt}=0=\frac{H}{c^2}{\bi u}-{\bi P}-\frac{1}{c^2}\frac{d{\bi S}}{dt}\times{\bi u}-\frac{1}{c^2}{\bi S}\times\frac{d{\bi u}}{dt}.
\label{eq:dKdt}
\end{equation}
If in (\ref{eq:dKdt}) we replace $d{\bi S}/dt={\bi P}\times{\bi u}$, and make the cross product with $d{\bi u}/dt$, we get
\begin{equation}
{\bi S}=\left(\frac{H-{\bi u}\cdot{\bi P}}{({d{\bi u}}/{dt})^2}\right)\left(\frac{d{\bi u}}{dt}\times{\bi u}\right)=\left(\frac{H-{\bi u}\cdot{\bi P}}{a^2}\right)({\bi a}\times{\bi u}),
\label{sdudtu}
\end{equation}
where we see that the spin ${\bi S}$ is orthogonal to the velocity and acceleration of ${\bi r}$.
From (\ref{eq:k}) ${\bi q}={\bi r}-{\bi S}\times{\bi u}/H$, by substituting this expression of ${\bi S}$ we finally obtain the definition of the position of the CM:
\begin{equation}
{\bi q}={\bi r}+\left(\frac{c^2-{\bi v}\cdot{\bi u}}{({d{\bi u}}/{dt})^2}\right)\,\frac{d{\bi u}}{dt}={\bi r}+\left(\frac{c^2-{\bi v}\cdot{\bi u}}{a^2}\right)\,{\bi a}.
\label{CMposicion}
\end{equation}
If we take ${\bi r}$ to the left hand side and produce the squared of this expression we get:
\[
({\bi q}-{\bi r})^2=\frac{(c^2-{\bi v}\cdot{\bi u})^2}{a^2},\quad \frac{c^2-{\bi v}\cdot{\bi u}}{a^2}=\frac{({\bi q}-{\bi r})^2}{c^2-{\bi v}\cdot{\bi u}},\quad a^2=\frac{(c^2-{\bi v}\cdot{\bi u})^2}{({\bi q}-{\bi r})^2},
\]
because $c^2-{\bi v}\cdot{\bi u}>0$. Therefore, equation (\ref{CMposicion}) can be rewritten as
\begin{equation}
{\bi a}=\frac{d{\bi u}}{dt}=\frac{c^2-{\bi v}\cdot{\bi u}}{({\bi q}-{\bi r})^2}({\bi q}-{\bi r}),
\label{eq:CCequ}
\end{equation}
and represents the acceleration of the CC in terms of the position and velocity of both points.

The expressions of the center of mass ${\bi q}$ and of the spins ${\bi S}$ and ${\bi S}_{CM}$, have been obtained without the explicit knowledge of the free Lagrangian $L_0$, in terms of the positions and velocities of both points, the CC and the CM.

We must remark that in the quantum Dirac's analysis, the only variable that defines the position of the electron ${\bi r}$, is contained in Dirac's spinor $\psi(t,{\bi r})$,  it is moving at the speed $c$ and it is also the point where the external electromagnetic field $A_\mu(t,{\bi r})$ is defined. Clearly, the spin with respect to the CC, ${\bi S}$, represents the classical equivalent of Dirac's spin operator.

\subsection{Dynamical equation in the center of mass frame}
The center of mass frame is defined as that frame where ${\bi P}=0$ and ${\bi q}=0$, and the CM is at rest ${\bi v}=0$, and located at the origin. From (\ref{eq:spinDyn}) we see that ${\bi S}$ is a constant vector in this frame. The equation (\ref{eq:espinS}) written in the center of mass frame looks
\begin{equation}
{\bi S}=-m{\bi r}\times{\bi u}.
\label{eq:Sxur}
\end{equation}
\begin{figure}[!hbtp]\centering%
\includegraphics[width=5.5cm]{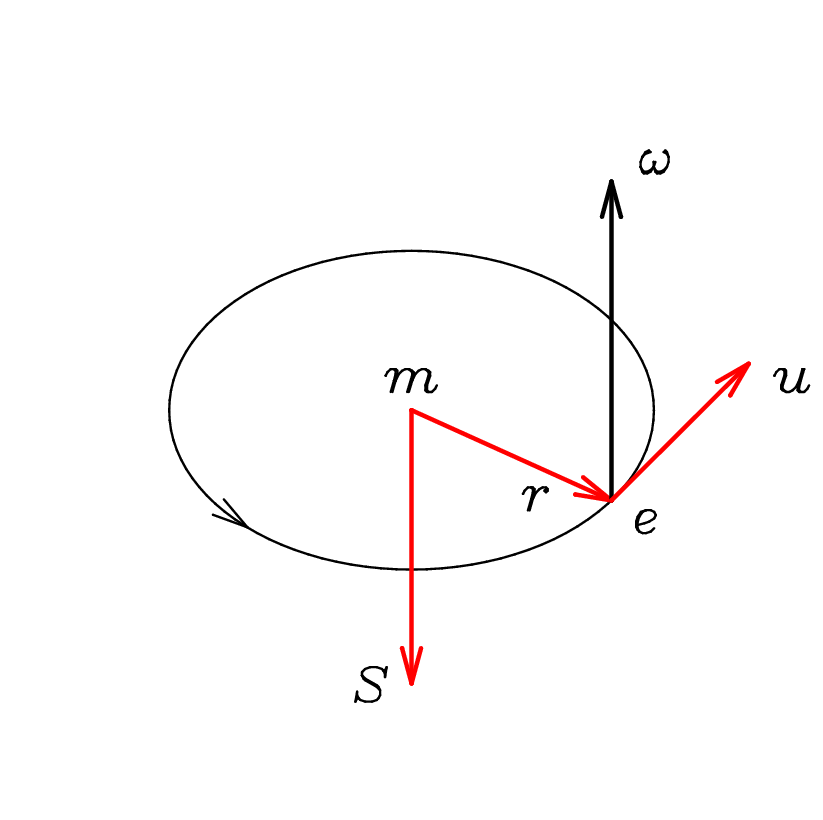}
\caption{This model represents the circular motion, at the speed of light, of the center
of charge of the Dirac particle in the center of mass frame, as described by the dynamical equation (\ref{eq:Sxur}). The center of mass is always a different point than the center of charge. The spin ${\bi S}$ has the opposite direction to the angular velocity in this frame $\bomega$. The trajectory is flat and the angular velocity has no component along the velocity ${\bi u}$. In general, for an arbitray inertial observer the spin ${\bi S}$ and the angular velocity $\bomega$ are not collinear vectors. The angular velocity has another component along the velocity $\bomega_{u}$ which produces the torsion of the CC motion, and which vanishes if the CC trajectory is flat. When we quantize this model the parameter $S=\hbar/2$.
The radius of this motion is $R_0=\hbar/2mc$, and the angular velocity is $\omega_0=2mc^2/\hbar$ in this frame. If the CM is moving this internal frequency decreases. The moving clock goes slowly.} 
\label{fig1:elecCM}
\end{figure}
It represents the dynamical equation of the point ${\bi r}$ of the free particle in the center of mass frame. If we take the time derivative of this expression we get ${\bi r}\times{\bi a}=0$ and, the vector ${\bi r}$ is collinear with the acceleration ${\bi a}$. The acceleration is orthogonal to the velocity ${\bi u}$ and thus ${\bi r}\cdot{\bi u}=d(r^2/2)/dt=0$, $r$ is constant and the motion of point ${\bi r}$ describes a circle of radius $R_0$ at the speed $c$, orthogonal to the constant spin direction.
The solution of this equation is depicted in the figure {\bf\ref{fig1:elecCM}}. We have also depicted de angular velocity $\bomega_0$ of the comoving frame attached to the point ${\bi r}$. The spin ${\bi S}$ and the angular velocity in this frame $\bomega_0$, have opposite directions. The constant parameter $S=mcR_0$.

\subsection{The angular velocity}
\label{angular}

The angular velocity corresponds to the rotation of the comoving Cartesian frame linked to the point ${\bi r}$. But this frame is completely arbitrary. Once the CC is moving, its dynamics defines the acceleration ${\bi a}$, orthogonal to the velocity ${\bi u}$, and therefore
these two orthogonal vectors with the vector ${\bi u}\times{\bi a}$ define an orthogonal comoving frame (the Frenet-Serret frame). Since ${\bi u}$ is a vector of constant absolute value, its time derivative the acceleration ${\bi a}$, is orthogonal to it and it can be written as
\[
{\bi a}={\bomega}\times{\bi u}.
\]
The cross product of this expression with ${\bi u}$ gives:
\[
{\bi a}\times{\bi u}={\bi u}({\bomega}\cdot{\bi u})-c^2\bomega.
\]
The first term is the component of the angular velocity along the velocity ${\bi u}$, $\bomega_u$ times $c^2$, so that $\bomega-\bomega_u=\bomega_p=({\bi u}\times{\bi a})/c^2$. Then the perpendicular component of the angular velocity $\bomega_p$ to the plane defined by the vectors ${\bi u}$ and ${\bi a}$, can be expressed as
\begin{equation}
\bomega_p=\frac{1}{c^2}({\bi u}\times{\bi a})=\frac{1}{c^2}\left(\frac{d{\bi r}}{dt}\times{\frac{d^2{\bi r}}{dt^2}}\right).
\label{omegaperp}
\end{equation}
This perpendicular component has the opposite direction than the CC spin (\ref{sdudtu}). In general the CM spin and the CC spin are not along the angular velocity $\bomega$. This perpendicular component is responsible for the curvature of the trajectory, while the $\bomega_u$ is related to the torsion of the trajectory. In the center of mass frame the trajectory of the point ${\bi r}$ is flat, there is no torsion and thus $\bomega_u=0$.

The component $\bomega_u$ produces the torsion of the trajectory and depends on the third orden derivative of the vector ${\bi r}$. This component of the angular velocity is
\begin{equation}
\bomega_u=\frac{1}{c^2a^2}\left(({\bi u}\times{\bi a})\cdot\frac{d{\bi a}}{dt}\right){\bi u}=\frac{1}{c^2a^2}\left[\left(\frac{d{\bi r}}{dt}\times{\frac{d^2{\bi r}}{dt^2}}\right)\cdot\frac{d^3{\bi r}}{dt^3}\right]\frac{d{\bi r}}{dt}.
\label{omegau}
\end{equation}

The evolution of the angular velocity is completely determined by the motion of the CC. The evolution of the CC determines both the trajectories of the CC and CM and the rotation of the body frame (Frenet-Serret frame) attached to the point ${\bi r}$.

We can find an alternative expression for the angular velocity $\bomega_u$ if we use the dynamical equation (\ref{eq:CCequ}). Taking the next order time derivative we have:
\[
\frac{d^3{\bi r}}{dt^3}=\frac{(-{\bi a}_{CM}\cdot{\bi u}-{\bi v}\cdot{\bi a})({\bi q}-{\bi r})^2-2({\bi q}-{\bi r})\cdot({\bi v}-{\bi u})(c^2-{\bi v}\cdot{\bi u})}{({\bi q}-{\bi r})^4}\,({\bi q}-{\bi r})+
\]
\[
+\frac{c^2-{\bi v}\cdot{\bi u}}{({\bi q}-{\bi r})^2}\,({\bi v}-{\bi u}).
\]
The term in squared brackets in (\ref{omegau})
\[
\left(\frac{d{\bi r}}{dt}\times{\frac{d^2{\bi r}}{dt^2}}\right)\cdot\frac{d^3{\bi r}}{dt^3}=\left(\frac{d^2{\bi r}}{dt^2}\times{\frac{d^3{\bi r}}{dt^3}}\right)\cdot\frac{d{\bi r}}{dt}
=\left({\bi a}\times{\frac{d^3{\bi r}}{dt^3}}\right)\cdot{\bi u},
\]
and since from (\ref{eq:CCequ}) ${\bi a}\simeq ({\bi q}-{\bi r})$, the last term in this expression is 
\[
\frac{c^2-{\bi v}\cdot{\bi u}}{({\bi q}-{\bi r})^2}\,({\bi a}\times({\bi v}-{\bi u}))\cdot{\bi u}.
\]
Similarly the term
\[
({\bi a}\times({\bi v}-{\bi u}))\cdot{\bi u}=(({\bi v}-{\bi u})\times{\bi u})\cdot{\bi a}=({\bi v}\times{\bi u})\cdot{\bi a}=({\bi u}\times{\bi a})\cdot{\bi v}.
\]
Collecting terms, instead of the expression (\ref{omegau}) we obtain
\begin{equation}
\bomega_u=\frac{1}{c^2a^2}\frac{c^2-{\bi v}\cdot{\bi u}}{({\bi q}-{\bi r})^2}\left[({\bi u}\times{\bi a})\cdot{\bi v}\right]{\bi u}=\left[\frac{\bomega_p\cdot{\bi v}}{c^2-{\bi v}\cdot{\bi u}}\right]{\bi u}.
\label{omegaunew}
\end{equation}
If the CM velocity ${\bi v}=0$, or along to the zitterbewegung plane, $\bomega_p\cdot{\bi v}=0$ and thus $\bomega_u=0$, the CC trajectory is a flat curve and there is no torsion. If $\bomega_p$ and ${\bi v}$ form an angle $\alpha<\pi/2$, $\bomega_u$ has the direction of ${\bi u}$ and the opposite direction if $\alpha>\pi/2$.

\begin{figure}[!hbtp]\centering%
\includegraphics[width=9cm]{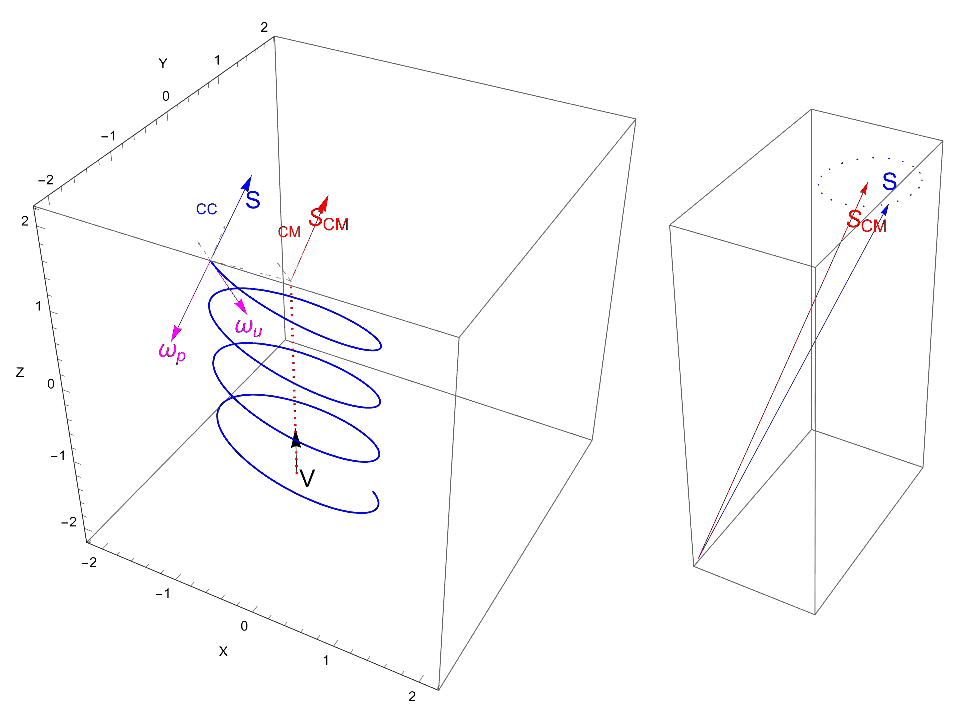}
\caption{Free motion of the Dirac particle with a constant CM velocity $v_z=0.12$ along OZ axis, and the CM spin orientation is $\theta=30^\circ$ and $\phi=65^\circ$. We see the zitter motion of the CC (blue) and the CM trajectory (dotted red). Attached to these points it is drawn the comoving cartesian frames. It is also depicted the center of mass spin ${\bi S}_{CM}$ (red) and the center of charge spin ${\bi S}$ (blue). On the right picture both spins are depicted to see that the ${\bi S}_{CM}$ is conserved but ${\bi S}$ is not, as given in (\ref{eq:spinDyn}) because the spin dynamics $d{\bi S}/dt={\bi p}\times{\bi u}$, is orthogonal to the constant linear momentum along the velocity ${\bi v}$, and precesses around ${\bi S}_{CM}$.\\
The two components of the angular velocity $\bomega_p$ and $\bomega_u$ (magenta), are also drawn. The component $\bomega_u$ has the opposite direction to the velocity of the CC ${\bi u}$, because the angle $\alpha$ between the vector $\bomega_p$ and the velocity ${\bi v}$ is $\alpha >\pi/2$, formula (24). If $\alpha<\pi/2$, $\bomega_u$ is in the same direction than the CC velocity ${\bi u}$ (see figure {\bf\ref{angular170}}). This component of the angular velocity is different from zero to produce the torsion of the CC trajectory and vanishes when this trajectory is flat. } 
\label{omegas}
\end{figure}

In the figure {\bf\ref{omegas}} we represent the motion of the free Dirac particle when integrating the dynamical equations with a {\tt Mathematica} notebook \cite{omegas}.

\begin{figure}[!hbtp]\centering%
\includegraphics[width=8cm]{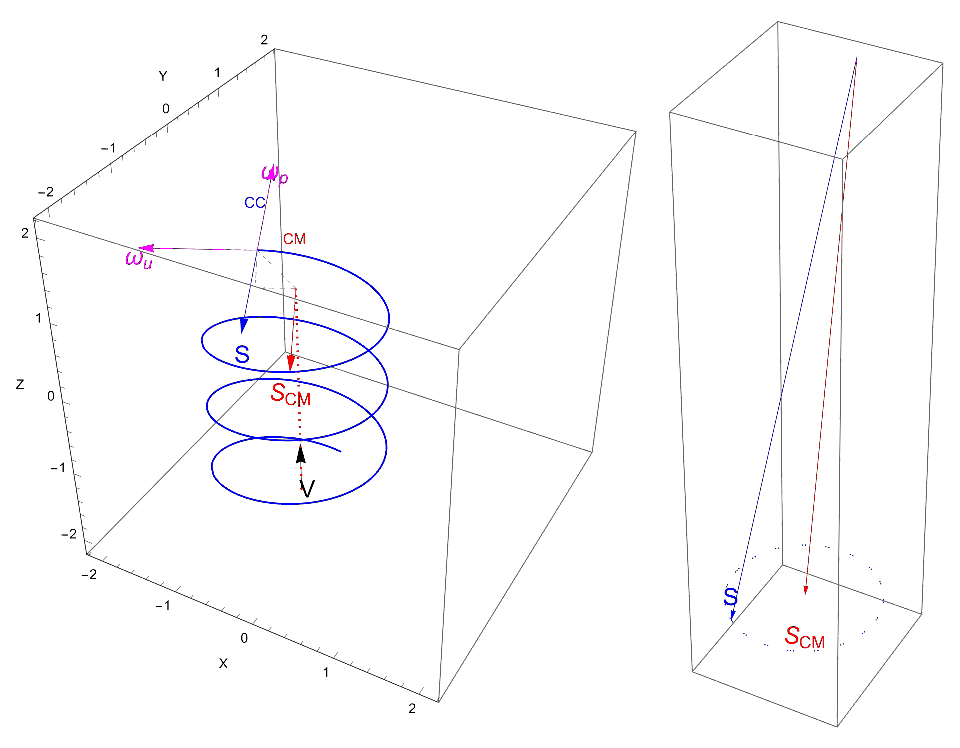}
\caption{Free motion of the Dirac particle with a constant CM velocity $v_z=0.14$ along OZ axis, and the CM spin orientation is $\theta=170^\circ$ and $\phi=270^\circ$. We see the zitter motion of the CC (blue) and the CM trajectory (dotted red). On the right picture both spins are depicted to see that the ${\bi S}_{CM}$ is conserved but ${\bi S}$ is not, as given in (\ref{eq:spinDyn}) because the spin dynamics $d{\bi S}/dt={\bi p}\times{\bi u}$, is orthogonal to the constant linear momentum along the velocity ${\bi v}$, and precesses around ${\bi S}_{CM}$.\\
The two components of the angular velocity $\bomega_p$ and $\bomega_u$ (magenta), are also drawn. The component $\bomega_u$ has in this case the same direction than the velocity of the CC ${\bi u}$, because the angle $\alpha$ between the vector $\bomega_p$ and the velocity ${\bi v}$, formula (24), is $\alpha<\pi/2$. } 
\label{angular170}
\end{figure}
 \newpage
\subsection{Dirac Hamiltonian}
\label{Hamiltonian}

If we take the scalar product of the expression (\ref{eq:dKdt}) with the velocity ${\bi u}$ we get
\begin{equation}
H={\bi P}\cdot{\bi u}+\frac{1}{c^2}{\bi S}\cdot\left(\frac{d{\bi u}}{dt}\times{\bi u}\right)={\bi P}\cdot{\bi u}-{\bi S}\cdot{\bomega}.
\label{eq:DiracH}
\end{equation}
The last term is in fact $-{\bi S}\cdot{\bomega}_p>0$, according to (\ref{sdudtu}) and (\ref{omegaperp}). The spin ${\bi S}$ is orthogonal to the velocity ${\bi u}$, and therefore to $\bomega_u$, and this term can be rewritten as $-{\bi S}\cdot{\bomega}>0$. The energy contains two terms: the first term, proportional to the linear momentum, represents the translational energy, while the second, proportional to the spin and angular velocity and to the motion of the CC (zitterbewegung), and independent of the CM velocity, represents the rotational energy $H_{rot}$ which is positive definite, because the spin ${\bi S}$ has the opposite direction to the angular velocity $\bomega_p$, and never vanishes. If we introduce in (\ref{eq:DiracH}) the expressions of ${\bi P}$ in (\ref{defHyp}) and ${\bi S}$ in (\ref{eq:espinS}),
\[
H=\gamma(v)m{\bi v}\cdot{\bi u}+\gamma(v)m\left[({\bi r}-{\bi q})\times{\bi u}\right]\cdot\left[\frac{1}{c^2}{\bi u}\times{\bi a}\right],
\]
we get $H=\gamma(v)mc^2$. It is this linear expression (\ref{eq:DiracH}) in terms of $H$ and ${\bi P}$, which represents the classical equivalent of Dirac's Hamiltonian \cite{RivasDirac}.

The rotational energy $H_{rot}=H-{\bi P}\cdot{\bi u}$, so that the ratio to the total energy is
\[
\frac{H_{rot}}{H}=\frac{c^2-{\bi v}\cdot{\bi u}}{c^2}=1-\frac{{\bi v}\cdot{\bi u}}{c^2}, 
\]
so that at low CM velocity the energy is basically rotational, while at high $v$ velocity ${\bi v}\cdot{\bi u}\simeq v^2$, and behaves like $H_{rot}/H\approx 1/\gamma(v)^2$, and translation energy dominates.

In the Lagrangian description we started with a mechanical system of 6 degrees of freedom. Three represent the position of a point ${\bi r}$, and another three $\balpha\in SO(3)$, that describe the orientation of a comoving frame attached to the point ${\bi r}$. The dynamics has established a constraint such that the degrees of freedom $\balpha$ can be selected to be the Frenet-Serret frame of the motion of ${\bi r}$, that satisfies a system of fourth-order differential equations. This point and their derivatives, completely determine the definition of any other observable, including the rotation of the comoving frame. The particle description has been reduced to the evolution of a single point, the center of charge.

Although the formalism is based on the existence of a Lagrangian, all the above expressions of the different observables $H$, ${\bi P}$, ${\bi q}$, ${\bi S}$, $\bomega=\bomega_p+\bomega_u$ and ${\bi S}_{CM}$,  for the free Dirac particle, have been obtained without the explicit knowledge of the free Lagrangian, by using Noether's theorem, and they are written  in terms of the position of the center of charge, and their time derivatives.

\subsection{Quantization in the CM frame}

The mechanical system depicted in the figure {\bf\ref{fig1:elecCM}}, once the spin direction is fixed along the $OZ$ axis, represents a three degree of freedom system: Two are the $x$ and $y$ coordinates of the point ${\bi r}$ on the $XOY$ plane and the third is the phase of the comoving frame attached to the point ${\bi r}$. But this phase is the same as the phase of the circular motion of constant radius $R_0$. This means that once the coordinate $x$ is given, the coordinate $y$ is determined, so that we are left with a mechanical system of just one-degree of freedom: the Cartesian coordinate $x$. This coordinate $x$ describes a harmonic motion of angular frequency $\omega_0$ in this frame. The energy of the particle at rest ${\bi v}=0$, is $H_0=mc^2$ and according to (\ref{eq:DiracH}) because ${\bi P}=0$, takes the constant value $H_0=-{\bi S}\cdot{\bomega}=S\omega_0$. According to the Atomic Principle this elementary particle has no excited states and therefore this motion represents a harmonic motion in its ground state. The quantized energy of this harmonic motion ground state is $\hbar\omega_0/2$.  The value of the spin in the center of mass frame is $S=\hbar/2$. This particle is a fermion. The angular velocity is thus $\omega_0=2mc^2/\hbar$, and this object has an internal frequency $\nu_0=2mc^2/h$, twice the frequency postulated by de Broglie.
A classical Dirac particle always has a unique internal frequency associated to its internal motion.
In the CM frame the Dirac particle is reduced to a one-dimensional harmonic oscillator. If the particle moves ${\bi v}\neq0$, the frequency decreases as $\nu(v)=\nu_0/\gamma(v)$. The local clock is going at a slower rate when moving \cite{RivasDirac}. See more details concerning time dilation in Section {\ref{interaction}}.

\section{The interaction Lagrangian}
\label{InterLag}
Although we do not know the free Lagrangian $L_0$, it will depend of the two invariant parameters $m$ and $S=\hbar/2$. Let us assume that the general Lagrangian under an interaction is $L$. Then the Lagrangians in the $\tau$ evolution description $\widetilde{L}$
and $\widetilde{L}_0$ are homogeneous functions of first degree in terms of the derivatives of the
boundary variables $x$. Therefore $\widetilde{L}-\widetilde{L}_0=\widetilde{L}_I$ is also a homogeneous function of the $\dot{x}$. This function $\widetilde{L}_I$ can also be expanded in the way (\ref{homogeneity}) where we include some interacting intensity factor $e$, the interacting charge of the particle:
\[
\widetilde{L}_I=e\left(A_0\dot{t}+{\bi A}\cdot\dot{\bi r}+{\bi C}\cdot\dot{\bi u}+{\bi D}\cdot\dot{\bomega} \right),
\]
where $eA_0=\partial\widetilde{L}_I/\partial\dot{t}$, $eA_i=\partial\widetilde{L}_I/\partial\dot{r}_i$,
$eC_i=\partial\widetilde{L}_I/\partial\dot{u}_i$, and $eD_i=\partial\widetilde{L}_I/\partial\omega_i$, are functions of the variables $x$ and $\dot{x}$. 
The general form of the Lagrangian $L$ of an interacting particle is written as $L=L_0+L_I$, the sum of the free Lagrangian $L_0$, plus the Lagrangian $L_I$, which we call the interaction Lagrangian.
The free Lagrangian is a function of the invariant parameters $m$ and $S=\hbar/2$, while the interaction Lagrangian will depend on the interacting parameter $e$ of the particle, that does not appear in the free $L_0$. We shall see in Section {\bf\ref{sec:invSp}} that $L_I$ will be independent of the acceleration ${\bi a}$ of the CC and of the angular velocity $\bomega$ and will be a function of only $t$, ${\bi r}$ and ${\bi u}$.

\section{Poincar\'e invariant magnitudes}
\label{PoinInvariant}
The 10 Noether constants of the motion (\ref{eq:n71}-\ref{eq:n7}) define two four-momenta. In terms of the constants of the motion (\ref{eq:n71}) and (\ref{eq:n72}), the energy-momentum four vector 
\begin{equation}
p^\mu\equiv(H/c,{\bi P}),
\label{pmu}
\end{equation}
and the Pauli-Lubanski four-momentum $w^\mu$ defined as
\begin{equation}
w^\mu=\frac{1}{2}\epsilon^{\mu\nu\sigma\lambda}p_\nu J_{\sigma\lambda},\qquad \epsilon^{0123}=+1.
\label{Wmu}
\end{equation}
They are orthogonal to each other $w^\mu p_\mu=0$.
The generalized angular momentum tensor $J^{\mu\nu}$ is the antisymmetric tensor 
\[
J^{\mu\nu}=x^\nu p^\mu-x^\mu p^\nu+S^{\mu\nu}=-J^{\nu\mu},
\]
which contains two parts: the space-time part
\[
L^{\mu\nu}=x^\nu p^\mu-x^\mu p^\nu=-L^{\nu\mu},
\]
which is not invariant under space-time translations, and the space-time translation invariant $S^{\mu\nu}=-S^{\nu\mu}$,
related to the CC spin of the system. The essential components of this are:
\[
S^{0i}=-({\bi S}\times{\bi u}/c)_i=-S^{i0}, \; i=1,2,3,\quad S_{ij}=-\frac{1}{2}\epsilon_{ijk}S^k,\;i,j,k=1,2,3.
\]
In matrix form
\[
S^{\mu\nu}=\pmatrix{0&(S_zu_y-S_yu_z)/c&(S_xu_z-S_zu_x)/c&(S_yu_x-S_xu_y)/c\cr
(S_yu_z-S_zu_y)/c&0&-S_z&S_y\cr(S_zu_x-S_xu_z)/c&S_z&0&-S_x\cr (S_xu_y-S_yu_x)/c&-S_y&S_x&0\cr}
\]
With this identification $J^{0i}=-J_{0i}=cK_i$ and $J_{ij}=-\frac{1}{2}\epsilon_{ijk}J^k$, where ${\bi K}$ and ${\bi J}$ are given in (\ref{eq:n73}) and (\ref{eq:n7}), respectively.
The time and space components of the $w^\mu$ are expressed in terms of all 10 Noether constants:
\begin{equation}
w^\mu\equiv({\bi P}\cdot{\bi J},H{\bi J}/c-c{\bi K}\times{\bi P})=({\bi P}\cdot{\bi S}_{CM},H{\bi S}_{CM}/c),
\label{wmu}
\end{equation}
after using (\ref{eq:J2}) ${\bi J}={\bi q}\times{\bi P}+{\bi S}_{CM}$ and (\ref{eq:K2}) ${\bi K}=H{\bi q}/c^2-{\bi P}t$.
The Pauli-Lubanski four-vector is finally written only in terms of the energy $H$, the linear momentum ${\bi P}$ and the CM spin ${\bi S}_{CM}$.

Under a general Lorentz transformation $x'^\mu=\Lambda^\mu_\nu x^\nu$, they transform as
\[
p'^\mu=\Lambda^\mu_\nu p^\nu,\quad w'^\mu=\Lambda^\mu_\nu w^\nu.
\]
The Poincar\'e invariant magnitudes
\[
p^\mu p_{\mu}=m^2c^2,\qquad w^\mu w_{\mu}=-m^2c^2S^2,
\]
define the mass of the particle $m$ and the absolute value of the spin in the center of mass frame $S$. In the free case they are also constants of the motion and therefore the corresponding constants take the same value in every inertial reference frame. They are the intrinsic properties of the elementary particle. Under any interaction, with the analysis that follows {\bf\ref{sec:invSp}} and {\bf\ref{sec:invofmass}} about the invariance of the spin and mass, these two intrinsic properties $m$ and $S$ are also invariant under any interaction if ${\bi P}$ represents the mechanical linear momentum ${\bi p}_m=\gamma(v)m{\bi v}$ and $H$ the mechanical energy $H_m=\gamma(v)mc^2$.

\subsection{Invariance of the spin}
\label{sec:invSp}
According to the Atomic Principle the interaction cannot modify the intrinsic parameters of the particle $m$ and $S$.
We must remark that the definition of the spin ${\bi S}={\bi u}\times{\bi U}+{\bi W}$ cannot be modified by the interaction so that the functions ${\bi U}$ and ${\bi W}$, which come from the dependence of $L_0$ of the acceleration and angular velocity, respectively, have to remain the same. The expression of the interaction Lagrangian $L_I$ has to be independent of the acceleration and angular velocity. Therefore the most general interaction Lagrangian in the time evolution description is
\begin{equation}
L_I=e\left(A_0(t,{\bi r},{\bi u})+{\bi A}(t,{\bi r},{\bi u})\cdot{\bi u}\right),
\label{interactionL}
\end{equation}
where the scalar function $A_0$ and the vector function ${\bi A}$ are, in general, functions of the variables $(t,{\bi r},{\bi u})$, but independent of ${\bi a}$ and $\bomega$. If we restrict these functions to be independent of ${\bi u}$ we obtain what is known as the {\it minimal coupling} of the electromagnetic interaction and $-A_0$ and ${\bi A}$ are, respectively, the scalar and vector potential of the electromagnetic field, $e$ is the electric charge of the particle and the interaction Lagrangian is a linear function of the velocity of the CC.

This classical analysis suggests that from the classical point of view more classical interactions can be described, provided we select the functions $A_0$ and ${\bi A}$, as functions also of the variables ${\bi u}$. This kind of analysis is still to be done.
In the analysis of the Poincar\'e invariant interaction between two Dirac particles \cite{2DiracParticles} the interacting term we found is not a linear function of the velocities of the two particles, and represents a non-electromagnetic interaction. It is a synchronous action at a distance and Poincar\'e invariant interaction, with velocity dependent terms.

\subsection{Invariance of the mass}
\label{sec:invofmass}

The mass $m$ is the other invariant property of the particle and is defined by
\[
H_m^2-c^2{\bi P}_m^2=m^2c^2,
\]
where the mechanical energy $H_m=\gamma(v)mc^2$ and the mechanical linear momentum ${\bi P}_m=\gamma(v)m{\bi v}$ are obtained from the free Lagrangian $L_0$. When the particle interacts it is described by the Lagrangian $\widetilde{L}=\widetilde{L}_0+\widetilde{L}_I$. The total energy $H$ and total linear momentum of the particle ${\bi P}$ under interaction will be defined by the same expressions (\ref{eq:n71}) and (\ref{eq:n72}), respectively. If we write the interaction Lagrangian in the form
\[
\widetilde{L}_I=-eA_0\dot{t}+e{\bi A}\cdot\dot{\bi r},
\]
according to the invariance of the spin $A_0$ and ${\bi A}$ are independent of the acceleration and angular velocity variables ${\bi a}$ and $\bomega$, respectively. If we also assume that they are independent of the velocity variables ${\bi u}$, like in the case of the electromagnetic interaction, from
$\widetilde{L}=\widetilde{L}_0+\widetilde{L}_I$, we obtain
\[
H=-T_0-\frac{d{\bi U}_0}{dt}-\frac{\partial{\widetilde{L}_I}}{\partial{\dot{t}}}=H_m+eA_0,\quad {\bi P}={\bi P}_m+\frac{\partial{\widetilde{L}_I}}{\partial{\dot{\bi r}}}={\bi P}_m+e{\bi A},
\]
and $eA_0$ could be interpreted as the transfer of energy and $e{\bi A}$ as the transfer of linear momentum by the electromagnetic interaction.

This implies that $H$ and ${\bi P}$ and the $A_0$ and ${\bi A}$ are not independent and satisfy the invariance of the mass equation:
\begin{equation}
H_m^2-c^2{\bi P}_m^2=(H-eA_0)^2-c^2({\bi P}-e{\bi A})^2=m^2c^4.
\label{eq:HyP}
\end{equation}
The functions $A_0$ and ${\bi A}$ are not uniquely defined. Let $\lambda(t,{\bi r})$ be an arbitrary differentiable function of $t$ and ${\bi r}$. If we replace
\[
eA'_0=eA_0-\frac{\partial \lambda}{\partial t}, \quad e{\bi A}'=e{\bi A}'+\nabla\lambda,
\]
the interaction Lagrangian 
\[
\widetilde{L}'_I=\widetilde{L}_I+\frac{d\lambda}{d\tau},
\]
so that the total Lagrangian changes with a total $\tau-$derivative, and the dynamical equations remain invariant. The interpretation of these functions $eA_0$ and $e{\bi A}$, which are not uniquely defined, as the energy and linear momentum transfer is doubtful.

	\subsection{Spin transformation}
\label{Spintransf}
If we call $w^\mu(0)\equiv(0,mc{\bi S}_{CM}(0))$ and $w^\mu({\bi v})\equiv(\gamma(v)m{\bi v}\cdot{\bi S}_{CM}({\bi v}),\gamma(v)mc{\bi S}_{CM}({\bi v}))$ to the Pauli-Lubanski four-vectors for the particle with the CM at rest and the CM moving at the velocity ${\bi v}$, respectively, the transformation $ w^\mu({\bi v})=\Lambda({\bi v})^\mu_\nu w^\nu(0)$ implies that
\begin{equation}
{\bi S}_{CM}({\bi v})=\frac{1}{\gamma(v)}{\bi S}_{CM}(0)+\frac{\gamma(v)}{(1+\gamma(v))c^2}({\bi v}\cdot{\bi S}_{CM}(0)){\bi v},
\label{eq:transfS}
\end{equation}
which represents how the CM spin transforms between the center of mass observer and an inertial or laboratory observer who sees the CM moving at the speed ${\bi v}$. 

The invariant property 
\[
w^\mu({\bi v})w_\mu({\bi v})=w^\mu(0)w_\mu(0)=\gamma(v)^2\left(({\bi v}\cdot{\bi S}_{CM}({\bi v}))^2-{\bi S}_{CM}({\bi v})^2\right)=-{\bi S}_{CM}(0)^2,
\]
and the relation between the absolute values of the spins in two different frames is:
\begin{equation}
S_{CM}(v,\phi)=S_{CM}(0)\sqrt{\frac{1-v^2}{1-v^2\cos^2\phi}},
\label{absolespinvar}
\end{equation}
where $\phi$ is the angle between the CM velocity ${\bi v}$ and the direction of the spin ${\bi S}_{CM}({\bi v})$.
In general $v$ and $\phi$, if the particle is not free, are functions of time in that frame, so that the absolute value of the spin is not a constant of the motion.

\begin{figure}[!hbtp]\centering%
 \includegraphics[width=8cm]{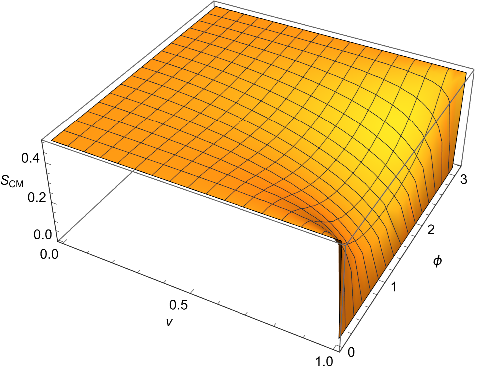}
\caption{Variation of the absolute value of the CM spin of (\ref{absolespinvar}) with the velocity of the CM and orientation $\phi$ between the velocity ${\bi v}$ and the center of mass spin ${\bi S}_{CM}({\bi v})$.}  
\label{variation}
\end{figure}

 The representation of this function is depicted in the figure {\bf\ref{variation}}.
The absolute value of the spin $S_{CM}({\bi v})$ is almost constant of value $1/2$ at low velocities but decreases at high velocity. If ${\bi v}$ and ${\bi S}_{CM}({\bi v})$ are parallel  or antiparallel vectors there is no change in the value of the spin $S_{CM}(v)=S_{CM}(0)=1/2$. If ${\bi v}$ and ${\bi S}_{CM}$ are orthogonal, $\phi'=\pi/2$, $\cos(\phi')=0$, and thus $S_{CM}(v)= 1/2\gamma(v)$ and decreases with $\gamma(v)$.  In the limit when $v\to c$,  $S_{CM}({\bi v})\to 0$.

If we send the particle from the center of mass frame with a velocity orthogonal to the spin, the transformed spin is also orthogonal to the velocity and of value $1/2\gamma(v)$. 
If we send the particle in the direction of the spin, the spin remains parallel or antiparallel to the velocity and the absolute value of the spin $1/2$ does not change. 
Except in the exceptional case that ${\bi v}$ and ${\bi S}_{CM}$ are parallel vectors the value of the spin, for an inertial observer who sees the particle at very high velocity, vanishes and the influence of the spin at high energy physics is almost irrelevant.

\section{Interaction of a Dirac particle with an electromagnetic field}
\label{interaction}
The Lagrangian of a Dirac particle in an external electromagnetic field is:
\[
L=L_0({\bi u},{\bi a},\bomega)+L_{em}(t,{\bi r},{\bi u}),\quad L_{em}=-eA_0(t,{\bi r})+e{\bi u}\cdot{\bi A}(t,{\bi r})
\]
where $L_0$ is the unknown free Lagrangian, $e$ is the electric charge of the particle and $A_0$ and ${\bi A}$, are, respectively, the external scalar and vector potentials defined at the CC of the particle and ${\bi u}$ is the CC velocity \cite{RivasDynamics}.

The free Lagrangian $L_0({\bi u},{\bi a},\bomega)$, is translation invariant and therefore it is a function of the velocity, acceleration of the CC, and of the angular velocity. The fourth-order Euler-Lagrange equations of the position variables are
\[
\frac{\partial L_0}{\partial{\bi r}}-\frac{d}{dt}\left(\frac{\partial L_0}{\partial{\bi u}}\right)+
\frac{d^2}{dt^2}\left(\frac{\partial L_0}{\partial{\bi a}}\right)+\frac{\partial L_{em}}{\partial{\bi r}}
-\frac{d}{dt}\left(\frac{\partial L_{em}}{\partial{\bi u}}\right)=0.
\]
Since $L_0$ is independent of ${\bi r}$, the above dynamical equation reads:
\[
-\frac{d}{dt}\left[\frac{\partial L_0}{\partial{\bi u}}-
\frac{d}{dt}\left(\frac{\partial L_0}{\partial{\bi a}}\right)\right]-e(\nabla A_0+\partial{\bi A}/\partial t)+e{\bi u}\times(\nabla\times{\bi A})=0,
\]
and the term between the squared brackets is ${\bi R}-d{\bi U}/dt$, and represents the mechanical linear momentum of the particle ${\bi p}_m=\gamma(v)m{\bi v}$, written in terms of the CM velocity. This equation is
\[
\frac{d{\bi p}_m}{dt}=e\left({\bi E}+{\bi u}\times{\bi B}\right),
\]
where ${\bi E}=-\nabla A_0-\partial{\bi A}/\partial t$, and ${\bi B}=\nabla\times{\bi A}$, are, respectively, the electric and magnetic external fields.
The CM position ${\bi q}$, is defined as ${\bi q}={\bi r}-{\bi S}\times{\bi u}/H_m$, in terms of the spin ${\bi S}$ and of the mechanical energy $H_m=\gamma(v)mc^2$, magnitudes which are not modified by the interaction, thus expression (\ref{CMposicion}) or its equivalent expression (\ref{eq:CCequ}) still holds. 

If we express the mechanical linear momentum ${\bi p}_m=\gamma(v)m{}{d{\bi q}/dt}$, we finally find the relativistic differential equations for the CC and CM positions of the Dirac particle in the presence of an external electromagnetic field ${\bi E}(t,{\bi r})$ and ${\bi B}(t,{\bi r})$. The  fields are defined at the CC position ${\bi r}$, in any arbitrary inertial reference frame \cite{RivasDynamics}: 
\begin{eqnarray}
\frac{d^2{\bi q}}{dt^2}&=&\frac{e}{m\gamma(v)}\left[{\bi E}+{\bi u}\times{\bi B}-\frac{1}{c^2}{\bi v}\left(\left[{\bi E}
+{\bi u}\times{\bi B}\right]\cdot{\bi v}\right)\right],\label{eq:d2qdt2}\\
\frac{d^2{\bi r}}{dt^2}&=&\frac{c^2-{\bi v}\cdot{\bi u}}{({\bi q}-{\bi r})^2}\left({\bi q}-{\bi r}\right),\label{eq:d2rdt2}
 \end{eqnarray}
with the constraint $|{\bi u}|=c$ and $|{\bi v}|<c$. 

Equation (\ref{eq:d2rdt2}) is equation (\ref{eq:CCequ}). Taking the second order derivative of ${\bi q}$ in (\ref{CMposicion}) and replacing it by (\ref{eq:d2qdt2}) we recover \cite{RivasDynamics} the fourth-order system of differential equations for the CC position ${\bi r}$, which is where the external fields ${\bi E}(t,{\bi r})$ and ${\bi B}(t,{\bi r})$, are defined. 
The $\gamma(v)$ factor in (\ref{eq:d2qdt2}) prevents that the velocity ${\bi v}$ reaches the speed of light, since the acceleration of the CM goes to zero when $v$ increases. An electron accelerated at the Tevatron to an energy of 1 TeV acquires a CM velocity $v/c=0.999\,999\,999\,999\,8694$ (twelve nines!) $\gamma(v)=1.58\cdot10^6$, and therefore at high energy it behaves as a free particle.
Since at any time $v<c$, the factor in (\ref{eq:d2rdt2})  $c^2-{\bi v}\cdot{\bi u}>0$, forbids that the acceleration ${\bi a}$ of the CC vanishes. The differential equation (\ref{eq:d2rdt2}) is not singular and thus being the acceleration ${\bi a}$ orthogonal to ${\bi u}$ guarantees that always $u=c$. It represents a kind of a relativistic generalization of a central or harmonic motion of point ${\bi r}$ around a fixed point ${\bi q}$. In these dynamical equations it is not considered the possible radiation-reaction term due to the emission of radiation of an accelerated charged particle.
This anlysis is left to a future paper.

Although this formalism is not written in covariant form in terms of four-vectors, it is nevertheless an invariant formalism in the sense that the expressions of the dynamical equations and of the different observables are written in the same form in each inertial reference frame, where the time variable is the time coordinate in that frame.

\subsection{Zitterbewegung}
\label{sec:zitter}

\begin{figure}[!hbtp]\centering%
\includegraphics[width=5.5cm]{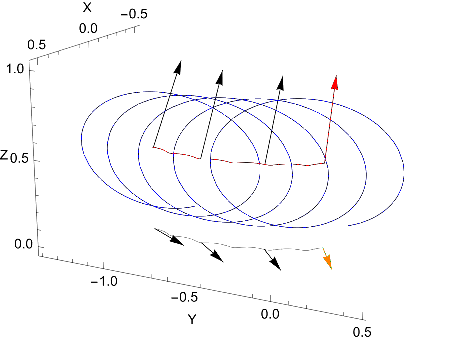}\includegraphics[width=5.5cm]{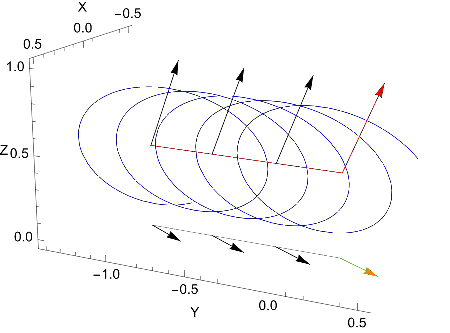}
\caption{Motion of the Dirac particle in a uniform magnetic field $B_z$. The initial velocity along OX axis is $v_x=0.05$, in natural units. We see the zitter motion of the CC (blue) and the cyclotron motion of the CM (red) which is almost a circle with a small oscillation of the same frequency as the CC motion, during a time of flight of around five turns of the CC. The CM velocity $|{\bi v}|$ is not a constant. It is also depicted the CM spin with orientation $\theta=30^\circ$, $\phi=60^\circ$, at different CM points. The red spin is the last value of the spin at the end of the integration time. In natural units the time of a turn is $\pi$ for the particle at rest and $\pi\gamma(v)$ for a moving Dirac particle. The center of mass spin precesses backwards with Larmor frequency, although the trajectory of the CM is a kind of a curly circle.\\
In the right figure we see the free motion of the particle, the zitteberwegung of the trajectory of the CC while the trajectory of the CM is a straight line with no zitter. It is also shown the projection of the CM trajectory and the ${\bi S}_{CM}$ projection on the XOY plane that remains constant.} 
\label{zitterQR}
\end{figure}

It was called by Schroedinger {\it zitterbewegung} to the trembling and high frequency motion of the electron. In this model it corresponds to the motion with torsion of the CC of the Dirac particle. Because the CM ${\bi q}$ is defined in terms of the CC by the expression (\ref{CMposicion}) then the trembling motion of the CC produces a small deviation of the CM trajectory without torsion in the free case, to a trajectory with some torsion under interaction, as depicted in the figure {\bf\ref{zitterQR}}, where we integrate the dynamical equations in a uniform magnetic field. We see that the CM trajectory has a small oscillation of the same frequency as the zitter motion of the CC, which implies a very small torsion of the CM trajectory. In the free case the trajectory of the CC is similar to this one but the trajectory of the CM is a straight line. In this example the CM spin has the orientation $\theta=30^\circ$, $\phi=60^\circ$.

\subsection{Time dilation}
\label{timedilation}
\begin{figure}[!hbtp]\centering%
\includegraphics[width=6cm]{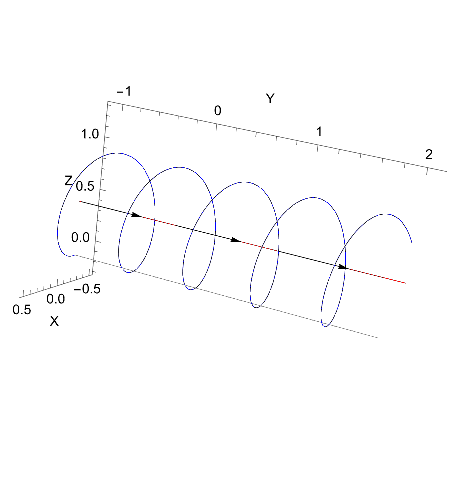}
\caption{Free motion of the Dirac particle with the CM spin along the OY axis and the CM velocity $v\equiv v_y=0.2$.
The transversal velocity of the CC is $u_\perp=\sqrt{c^2-v^2}$ and the time of a turn is $T(v)=\gamma(v)T_0$, where $T_0$ is the time of a turn when the CM is at rest. The local clock is going at a slower rate when moving.}
\label{timedilation}
\end{figure}

The time $T_0$ of the motion of the CC when the CM is at rest during a zitter turn is $T_0=2\pi R_0/c$. When the particle is moving with a velocity $v$ along the spin direction, like in the figure {\bf\ref{timedilation}} the velocity of the CC ${\bi u}$ has a longitudinal component $u_y=v$ and a transversal component $u_\perp=\sqrt{c^2-v^2}$. Now the time taken during a turn is $T(v)=2\pi R_0/u_\perp=\gamma(v)T_0$. What we measure is that the local clock of the moving particle is going at a slower rate when moving. It takes more time to complete a cycle. This time dilation is independent of the relative orientation between the CM spin and the initial CM velocity and it is the same as the time dilation between the laboratory observer and the inertial observer at rest with the CM of the particle. As we shall see in Section {\bf\ref{Natural}}, in natural units $R_0=1/2$ and $c=1$ and the time of a turn is $T_0=\pi$ in the center of mass frame and $T(v)=\gamma(v)\pi$ for an arbitrary observer who sees the CM moving at the speed $v$.

\subsection{Magnetic moment $\bmu_{CM}$ of the Dirac particle}

For the Dirac particle of electric charge $e$, the relationship between its magnetic moment with respect to the CM and the spin with respect to the CM \cite{magneticDirac} should include the factor $g=2$, the gyromagnetic ratio, and will be
\begin{equation} 
 \bmu_{CM}=\frac{e}{m\gamma(v)}{\bi S}_{CM}.
\label{eq:defMu}
\end{equation}  
We take, for the Dirac particle, as the exact relationship between the instantaneous magnetic moment and the spin with respect to the center of mass, the above definition (\ref{eq:defMu})
as predicted by Dirac equation. The $\gamma(v)$ factor implies that the magnetic moment of the Dirac particle decreases at very high energy.

Using the expression of ${\bi S}_{CM}$ in (\ref{eq:espinSCM}) the appropriate definition of the instantaneous magnetic moment with respect to the center of mass of the Dirac particle, in terms of the classical variables which describe its motion, will be:
\begin{equation}
{\bmu}_{CM}=-e({\bi r}-{\bi q})\times({\bi u}-{\bi v}).
\label{eq:mupointCM}
\end{equation}
The absolute value in the center of mass frame is
\[
\mu_B=\frac{e\hbar}{2m},
\]
called Bohr's magneton. This value is also the average value of the magnetic moment during a complete turn of the CC around the CM in the free case.

\subsection{Electric dipole moment ${\bi d}_{CM}$ of the Dirac particle}

The electric dipole moment of a point charge $e$ located at the point ${\bi r}$, with respect to the CM ${\bi q}$ is:
\begin{equation}
{\bi d}_{CM}=e({\bi r}-{\bi q}).
\label{elecdipolMom}
\end{equation}
In chapter 6 of \cite{Rivasbook} it is shown that the above classical definition of electric dipole moment (\ref{elecdipolMom}) leads, when quantizing this model, to Dirac's electric dipole moment operator. It is also shown in that chapter that the absolute value of this dipole gives rise to Darwin's term of Dirac's Hamiltonian $-{\bi d}_{CM}\cdot{\bi E}(t,{\bi r})$.

The absolute value of the electric dipole moment in the center of mass frame is $d=eR_0$. However the average value of the electric dipole moment during a complete turn of the CC around the CM is zero.

As predicted by Dirac's theory in the quantum analysis, the classical Dirac particle has also both electric dipole moment and magnetic dipole moment of value in natural units of $1/2$, as we shall see below.

\subsection{Experimental measurement of $\bmu_{CM}$ and ${\bi d}_{CM}$}
As analyzed in the work \cite{magneticDirac} one thing is the theoretical prediction of the expressions of $\bmu_{CM}$ and ${\bi d}_{CM}$ and another is the experimental measurement. To measure we have to interact with the particle and this modifies the motion of both points ${\bi r}$ and ${\bi q}$ and the velocities ${\bi u}$ and ${\bi v}$. If the measurement implies some average value of both magnitudes during a long period of time, we expect that the value of $\mu$ will be Bohr's magneton and the value of $d$ is $eR_0$ in the center of mass frame. But the experimental measurement is not done in the center of mass frame because the interaction modifies the CM velocity.

We are going to see in figure {\bf\ref{fig2:initial}} that the separation between ${\bi r}$ and ${\bi q}$ depends on the relative orientation $\alpha$ between the CM velocity ${\bi v}$ and the CM spin ${\bi S}_{CM}$, so that $|{\bi r}-{\bi q}|$ is in the range $[1/2-(v/2)\sin\alpha,1/2+(v/2)\sin\alpha]$ in natural units, and this separation is changing with the dynamics. Similarly the term $|{\bi u}-{\bi v}|$ is not constant along the trajectory, so that all factors which define both momenta are changing. In the mentioned article
\cite{magneticDirac} it is analyzed the average value of the above expressions during a finite number of internal turns of the motion of the CC around the CM for a Dirac particle in a Penning trap, to obtain a numerical average.
We obtain that $\mu\le\mu_B$ and that $d\ge eR_0$. 

\section{Natural units}
\label{Natural}
The system of differential equations (\ref{eq:d2qdt2}) and (\ref{eq:d2rdt2}) can be rewritten in terms of dimensionless variables, once the expressions of the fields are known.
If we replace ${\bi u}=c\widetilde{\bi u}$,
${\bi v}=c\widetilde{\bi v}$, ${\bi r}=2R_0\widetilde{\bi r}$,  ${\bi q}=2R_0\widetilde{\bi q}$ and $t=\tau_0\widetilde{t}$, where $R_0=\hbar/2mc$, is the separation between the CC and CM of the Dirac particle at rest and $\tau_0=2R_0/c$,
the equation (\ref{eq:d2rdt2}) becomes
\[
\frac{d^2{\bi r}}{dt^2}=\frac{c^2}{2R_0}\frac{d^2\widetilde{\bi r}}{d\widetilde{t}^2}=\frac{c^2}{2R_0}
\frac{1-{\widetilde{\bi v}}\cdot{\widetilde{\bi u}}}{({\widetilde{\bi q}}-{\widetilde{\bi r}})^2}({\widetilde{\bi q}}-{\widetilde{\bi r}})
\]
the coefficients cancel and the equation remains of the same form as in (\ref{eq:d2rdt2}) in terms of dimensionless variables and $c=1$.
For the other equation we need the explicit form of the fields. We shall analyze this in different examples.

We have defined two natural units: the speed of light $c$ and the natural unit of length $2R_0=\hbar/mc$, where $R_0$ represents the separation between the CC and the CM. This defines a natural unit of time $\tau_0=2R_0/c$, as the time taken by a light ray to cover the distance $2R_0$. The time of a complete trajectory of the CC around the CM at the center of mass frame is $\pi$. All boundary variables are dimensionless.
 
From the mechanical point of view we need an extra natural unit of mass. The usual assumption is to define as the natural unit of action such that Planck's constant $\hbar=1$. In this way since
\[
2R_0=\frac{\hbar}{mc}=1,\quad\Rightarrow\quad m=1\;{\rm n.u.},
\]
therefore the natural unit of mass is that unit where the mass of the electron is $1$. 
The energy of the particle is expressed in terms of the center of mass velocity as $H(v)=\gamma(v)mc^2$ which in natural units for the electron $H(v)=\gamma(v)$. The linear momentum is expressed as ${\bi p}(v)=\gamma(v){\bi v}$. The value of the spin in the center of mass frame $S=\hbar/2=1/2$ in natural units. The general expressions of both spins for the electron in natural units are
\[
{\bi S}=-{\gamma(v)}({\bi r}-{\bi q})\times{\bi u},
\]
\[
{\bi S}_{CM}=-{\gamma(v)}({\bi r}-{\bi q})\times({\bi u}-{\bi v}),
\]
where the variables ${\bi r}$, ${\bi q}$, ${\bi u}$ and ${\bi v}$ are expressed in natural units. In general, the absolute value of both spins will be a function of the center of mass velocity ${\bi v}$, $S(v)$ and $S_{CM}(v)$,
with the values at rest  $S(0)=S_{CM}(0)=1/2$, as described in Section {\bf\ref{Spintransf}}. The dynamics modifies the variables they depend and the final value will be determined during the dynamical process. 

From the electromagnetic point of view we need a natural unit for the electric charge. The usual assumption is to define the electric charge of the electron through the fine structure constant and this defines the permitivity of the vacuum $\epsilon_0$,
\[
\alpha=\frac{e^2}{4\pi\epsilon_0 c\hbar}=0.007297, \quad\rightarrow\quad e=1\,{\rm n.u.},\quad \frac{1}{4\pi\epsilon_0}=\alpha,
\]
The electron (and positron) has as intrinsic parameters $m=1$, $S=1/2$ and $e=\pm1$. 

What we have is to translate the international system of units to this natural system of units.
The relationship for the fundamental units of mass [M], length [L], time [T] and electric charge [Q] is:
\[
1\; {\rm n.u.} [M]=m_e=9.109534 \cdot10^{-31}{\rm Kg},\quad\rightarrow 1\;{\rm Kg}\equiv 1.09775\cdot10^{30}\;{\rm n.u.}
\]
\[
1\; {\rm n.u.} [L]=2R_0=\frac{\hbar}{mc}=3.86153\cdot10^{-13}{\rm m},\quad\rightarrow 1\; {\rm m}\equiv 2.58965\cdot10^{12}\;{\rm n.u.}
\]
\[
1\; {\rm n.u.} [T]=\tau_0=\frac{2R_0}{c}=6.44034\cdot10^{-22}\;{\rm s},\quad \rightarrow 1\; {\rm s}\equiv 7.76357\cdot10^{20}\;{\rm n.u.}
\]
\[
e=1 \;{\rm n.u.}\; [Q]=1.6021892\cdot10^{-19} {\rm C},\quad\rightarrow 1 \;{\rm C}\equiv 6.24146\cdot10^{18}\;{\rm n.u.}
\]
The electric field in the International System of units is expressed in V/m. According to the equivalence among units
\begin{equation}
1 \;{\rm V/m}=1\; {\rm m\, Kg\, s^{-2}\,C^{-1}},\quad 1\; {\rm V/m}=7.55676\cdot 10^{-19}\;{\rm n.u.} 
\label{voltM}
\end{equation}
The magnetic field is expressed in teslas. Since
\begin{equation}
1 \;{\rm T}=1\; {\rm Kg\, s^{-1}\,C^{-1}},\quad 1\; {\rm T}=2.26546\cdot 10^{-10}\;{\rm n.u.} 
\label{Teslanu}
\end{equation}

\subsection{Boundary Conditions}

To integrate the fourth-order system of differential equations of the point ${\bi r}(t)$ we need to supply the 12 values of ${\bi r}(0)$, ${\bi r}^{(1)}(0)$, ${\bi r}^{(2)}(0)$ and ${\bi r}^{(3)}(0)$, of the initial values of the point and their derivatives up to the third order. Nevertheless since the velocity ${\bi u}\equiv{\bi r}^{(1)}$ is of constant value $u=c$ and it is orthogonal to the acceleration  ${\bi u}\cdot{\bi a}={\bi r}^{(1)}\cdot{\bi r}^{(2)}=0$, we are left with only 10 independent boundary conditions.

To integrate the system of differential equations (\ref{eq:d2qdt2}) and (\ref{eq:d2rdt2}), we have to establish the appropriate boundary conditions for the positions and velocities of both points, the CC and the CM of each particle.
These 12 boundary values for the variables ${\bi r}(0)$, ${\bi u}(0)$, ${\bi q}(0)$ and ${\bi v}(0)$, are finally to be expressed in terms of 10 essential parameters, because $|{\bi u}(0)|=1$ and ${\bi r}(0)-{\bi q}(0)$ is orthogonal to ${\bi u}(0)$. These 10 essential parameters are going to be those parameters that define the relationship between the center of mass observer of the particle and any arbitrary inertial observer or laboratory observer who sees the CM of the particle moving at the speed ${\bi v}$. 

If $t^*$ and ${\bi r}^*$ are the time and position of the CC of the particle for the center of mass observer $O^*$ of this particle, and $t$ and ${\bi r}$ are the time and position of the CC for any arbitrary inertial observer, they are related by the Poincar\'e transformation:
\[
x^{\mu}=\Lambda^\mu_\nu x^{*\nu}+a^\mu,\quad x^\mu\equiv(ct,{\bi r}),\quad x^{*\mu}\equiv(ct^*,{\bi r}^*), \quad a^\mu\equiv(cb,{\bi d}),
\]
and $\Lambda=L({\bi v})R(\psi,\theta,\phi)$ is a general Lorentz transformation, as a composition of a rotation $R$, followed by a boost of velocity ${\bi v}$, or pure Lorentz transformation $L({\bi v})$.

We are going to describe now the rotation $R(\psi,\theta,\phi)$, the boost or pure Lorentz transformation $L({\bi v})$ with arbitrary velocity ${\bi v}$, the space translation  $T({\bi d})$ of displacement ${\bi d}$, and finally the time translation $T(b)$, of value $b$. 

\begin{figure}[!hbtp]\centering%
\includegraphics[width=6cm]{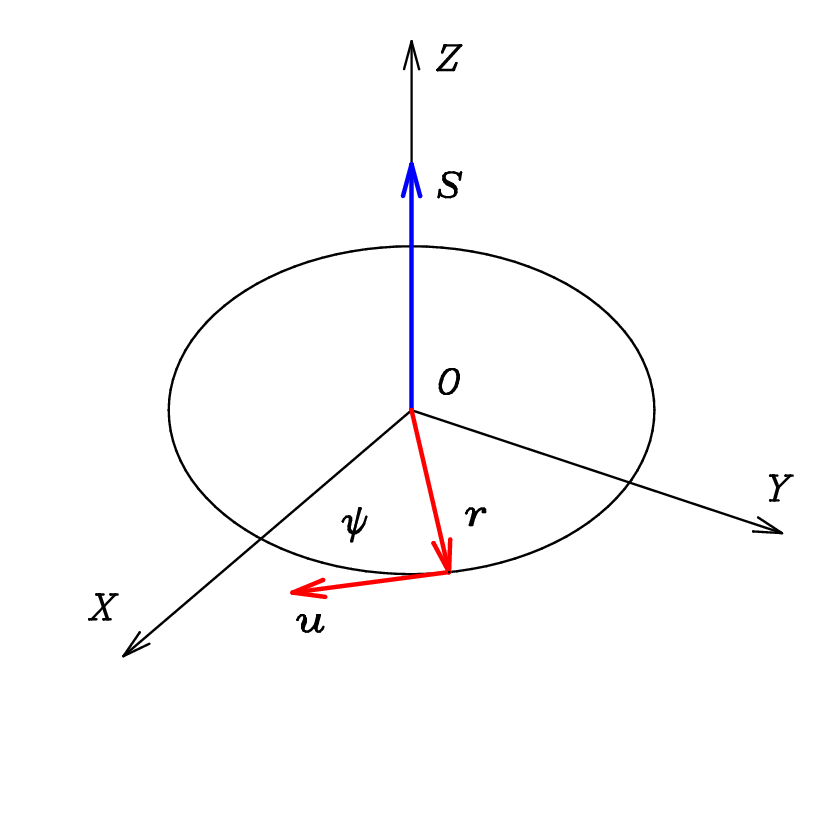}
\caption{Dirac particle in the center of mass reference frame, with the spin along the OZ axis. The initial position and velocity of the CC on the $XOY$ plane is determined by the phase  $\psi$. The radius of this motion is $R_0=1/2$, in natural units.} 
\label{fig2:initial}
\end{figure}

Let us first analyze the rotation. Let us assume we have the Dirac particle in the center of mass reference frame, as depicted in the figure {\bf\ref{fig2:initial}}, with the spin along $OZ$ axis.

If we initially have the CC at the position $(1/2,0,0)$ on the $OX$ axis, we rotate first an angle $\psi$, around the $OZ$ axis to modify the CC position. Next, we change the orientation of the spin by modifying the zenithal angle $\theta$, and azimuthal angle $\phi$, so that the arbitrary rotation is the composition of the rotations
\[
R(\psi,\theta,\phi)=R_{OZ}(\phi)R_{OY}(\theta)R_{OZ}(\psi),
\]
\[
R_{OZ}(\psi)=\pmatrix{\cos\psi&-\sin\psi&0\cr \sin\psi&\cos\psi&0\cr 0&0&1},
\]
\[
R_{OY}(\theta)=\pmatrix{\cos\theta&0&\sin\theta\cr 0&1&0\cr -\sin\theta&0&\cos\theta },\quad R_{OZ}(\phi)=\pmatrix{\cos\phi&-\sin\phi&0\cr \sin\phi&\cos\phi&0\cr 0&0&1}.
\]
If initially at time $t^*_0$, and in natural units the CC is at the point $(1/2,0,0)$, then the position, velocity and acceleration of the CC, before the rotation, since $c=1$ and $a=c^2/R_0=2$, are:
\[
{\bi r}^*(t^*_0)=\pmatrix{1/2\cr 0\cr 0},\quad {\bi u}^*(t^*_0)=\pmatrix{0\cr -1\cr 0}, \quad {\bi a}^*(t^*_0)=\pmatrix{-2\cr 0\cr 0}.
\]
These variables, after the rotation, become ${\bi r}^*_0=R_{OZ}(\phi)R_{OY}(\theta)R_{OZ}(\psi){\bi r}^*(t^*_0)$, and the same for ${\bi u}^*(t^*_0)$ and ${\bi a}^*(t^*_0)$, at the same time $t^*_0$, and thus:
 \begin{equation}
{\bi r}^*_0=\frac{1}{2}\pmatrix{\cos\theta\cos\phi\cos\psi-\sin\phi\sin\psi\cr \cos\theta\sin\phi\cos\psi+\cos\phi\sin\psi\cr
-\sin\theta\cos\psi},
 \label{eq:valoresinicialesr}
 \end{equation}
  \begin{equation}
  {\bi u}^*_0=\pmatrix{\cos\theta\cos\phi\sin\psi+\sin\phi\cos\psi\cr \cos\theta\sin\phi\sin\psi-\cos\phi\cos\psi\cr
-\sin\theta\sin\psi},
 \label{eq:valoresinicialesu}
 \end{equation}
\begin{equation}
{\bi a}^*_0=-2\pmatrix{\cos\theta\cos\phi\cos\psi-\sin\phi\sin\psi\cr \cos\theta\sin\phi\cos\psi+\cos\phi\sin\psi\cr
-\sin\theta\cos\psi}=-4{\bi r}^*_0.
 \label{eq:valoresinicialesac}
\end{equation}
The Lorentz transformation $L({\bi v})$ is given in (\ref{eq:Tdev})
where $\gamma\equiv(1-v_x^2-v_y^2-v_z^2)^{-1/2}$. Finally the two translations $T({\bi d})$ and $T(b)$.
The corresponding time $t_0$ and position ${\bi r}_0$ of the CC of the same event, for any arbitrary inertial observer in natural units are:
\begin{equation}
t_0=\gamma\left(t^*_0+{{\bi v}\cdot{\bi r}^*_0}\right)+b,\quad {\bi r}_0={\bi r}^*_0+\gamma{\bi v}t^*_0+\frac{\gamma^2}{1+\gamma}{({\bi v}\cdot{\bi r}^*_0){\bi v}}+{\bi d},
\label{eq:transt0r0}
\end{equation}
where $b$ is the time translation, ${\bi d}$ is the space translation in natural units and ${\bi v}$ is the velocity of the center of mass observer $O^*$ as measured by the observer $O$. It represents, therefore, the velocity of the center of mass of the particle for the arbitrary laboratory observer $O$.

If the initial instant to integrate the equations in the reference frame of the Laboratory $O$, is the time $t_0=0$, this corresponds to
$\gamma t^*_0=-\gamma {\bi v}\cdot{\bi r}^*_0-b$, for the center of mass observer of the particle, and therefore the initial position ${\bi r}_0$ of the CC for the laboratory observer at the initial instant  $t_0=0$, instead of (\ref{eq:transt0r0}) is:
\begin{equation}
{\bi r}_0={\bi r}^*_0-\frac{\gamma}{1+\gamma}{({\bi v}\cdot{\bi r}^*_0){\bi v}}-b{\bi v}+{\bi d}.
\label{ecuac:r0}
\end{equation}
For the remaning observables, 
 \begin{equation}
{\bi u}_0=\frac{{\bi u}^*_0+\gamma{\bi v}+\frac{\gamma^2}{(1+\gamma)}{({\bi v}\cdot{\bi u}^*_0){\bi v}}}{\gamma(1+{\bi v}\cdot{\bi u}^*_0)},
 \label{ecuac:u0}
 \end{equation} 
and for the acceleration we use the equation (\ref{eq:transaccel}) in natural units
 \begin{equation}
{\bi a}_0=\frac{(1+{\bi v}\cdot{\bi u}^*_0){\bi a}^*_0-({\bi v}\cdot{\bi a}^*_0){\bi u}^*_0-\frac{\gamma}{(1+\gamma)}({\bi v}\cdot{\bi a}^*_0){\bi v}}{\gamma^2(1+{\bi v}\cdot{\bi u}^*_0)^3},
 \label{ecuac:a0}
 \end{equation}
where ${\bi r}^*_0$, ${\bi u}^*_0$ and ${\bi a}^*_0$ are given above in (\ref{eq:valoresinicialesr}), (\ref{eq:valoresinicialesu}) and (\ref{eq:valoresinicialesac}), respectively.

With these values we have the initial boundary conditions for the variables ${\bi r}_0$, ${\bi u}_0$ and ${\bi v}_0={\bi v}$. What we have left is ${\bi q}_0$, the initial position of the CM in this reference frame.

The definition of the center of mass position of the Dirac particle in any reference frame at any time is given in (\ref{CMposicion})
\[
{\bi q}(t)={\bi r}(t)+\frac{1-{\bi v}(t)\cdot{\bi u}(t)}{{\bi a}(t)^2}\,{\bi a}(t),
\]
then the initial value of the CM ${\bi q}_0$, in the laboratory frame is:
\begin{equation}
{\bi q}_0={\bi r}_0+\frac{1-{\bi v}\cdot{\bi u}_0}{{{\bi a}_0}^2}\,{\bi a}_0,
\label{eq:iniPOSCM}
\end{equation}
in terms of the position, velocity and acceleration of the CC in this reference frame and of the velocity of the CM, all these variables defined at the initial time $t_0=0$.

Since ${\bi a}^*_0=-4{\bi r}^*_0$, taking the squared of (\ref{ecuac:a0}) we get
\[
{{\bi a}_0}^2=\frac{4}{\gamma^4(1+{\bi v}\cdot{\bi u}^*_0)^4},
\]
From (\ref{ecuac:u0}) the term
\[
1-{\bi v}\cdot{\bi u}_0=\frac{1}{\gamma^2(1+{\bi v}\cdot{\bi u}^*_0)}\;\Rightarrow\;
\frac{1-{\bi v}\cdot{\bi u}_0}{{\bi a}_0^2}{\bi a}_0=\frac{1}{4}{\gamma^2}(1+{\bi v}\cdot{\bi u}^*_0)^3{\bi a}_0,
\]
and if we substitute for ${\bi a}_0$ the expression (\ref{ecuac:a0}), where ${\bi a}^*_0=-4{\bi r}^*_0$, if we use (\ref{ecuac:r0}) the initial position (\ref{eq:iniPOSCM}) of the CM, ${\bi q}_0$, is
\begin{equation}
{\bi q}_0
={\bi v}\times({\bi u}^*_0\times{\bi r}^*_0)-b{\bi v}+{\bi d}.
\label{inicialcondCM}
\end{equation}

As a summary, the boundary conditions at $t_0=0$, in the laboratory frame, are:
 \begin{equation}
{\bi r}_0={\bi r}^*_0-\frac{\gamma}{1+\gamma}{({\bi v}\cdot{\bi r}^*_0){\bi v}}-b{\bi v}+{\bi d}, 
 \label{eq:initialcond1}
 \end{equation}
 \begin{equation}
 {\bi u}_0=\frac{{\bi u}^*_0+\gamma{\bi v}+\frac{\gamma^2}{(1+\gamma)}{({\bi v}\cdot{\bi u}^*_0){\bi v}}}{\gamma(1+{\bi v}\cdot{\bi u}^*_0)},
 \label{eq:initialcond3}
 \end{equation}
 \begin{equation}
{\bi q}_0={\bi v}\times({\bi u}^*_0\times{\bi r}^*_0)-b{\bi v}+{\bi d},
 \label{eq:initialcond2}
 \end{equation}
 \begin{equation} 
 {\bi v}_0={\bi v}.
  \label{eq:initialcond4}
 \end{equation}
 
\begin{figure}[!hbtp]\centering%
\includegraphics[width=7cm]{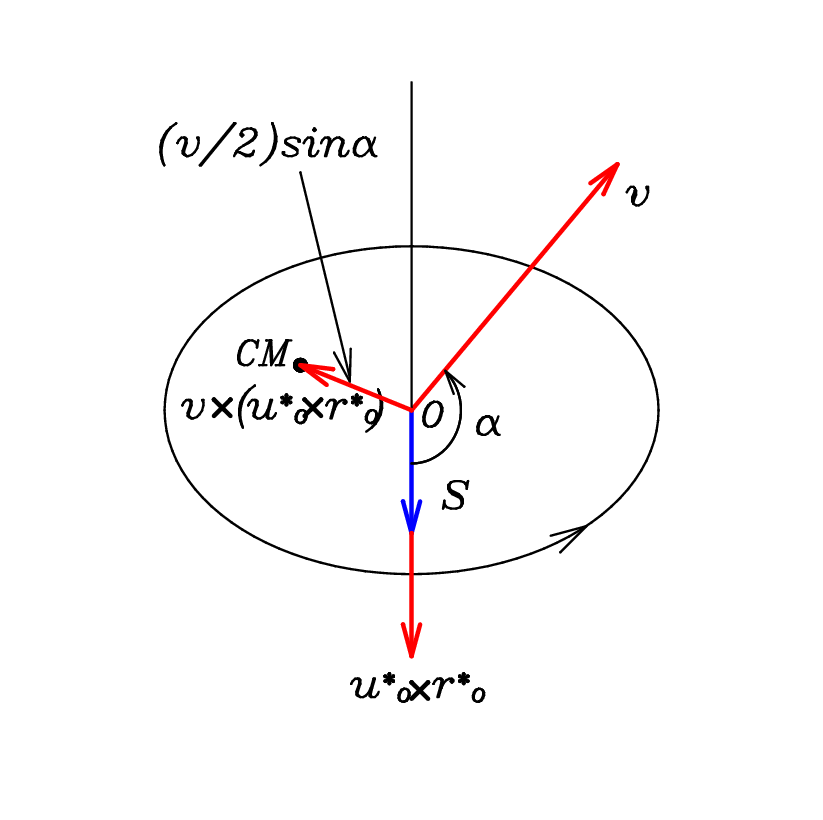}
\caption{Initial position of the CM when the velocity of the CM
${\bi v}$ and the spin direction form an angle $\alpha$. It is perpendicular to the vectors ${\bi v}$ and ${\bi u}^*_0\times{\bi r}^*_0$, of distance to the center $O-CM=(v/2)\sin\alpha$, in natural units, and is independent of the initial position and velocity of the CC. In general, the separation between both centers CC and CM, is in the interval $|{\bi r}-{\bi q}|\in[1/2-(v/2)\sin\alpha,1/2+(v/2)\sin\alpha]$, in natural units, so that this separation is not a constant of the motion.} 
\label{fig3:posCMini}
\end{figure}

These boundary conditions are expressed in terms of 10 essential parameters: the space and time parameters ${\bi d}$ and $b$ that define the initial location of the CC and CM and the initial instant of the integration, respectively. The three parameters ${\bi v}$ that define the initial velocity of the CM and finally the three angles $\psi$, the initial phase of the CC, and $\theta$ and $\phi$ that define the initial orientation of the spin in the center of mass frame.

In general, the initial position of the CM for the laboratory observer, given in (\ref{eq:initialcond2}), is contained in the zitterbewegung plane (with ${\bi d}=0=b$),
is independent of the initial position of the CC, ${\bi r}^*_0$,
and is depicted in the figure {\bf\ref{fig3:posCMini}}.
The distance to the center is $(v/2)\sin\alpha$.
If ${\bi v}$ is perpendicular to the ziterbewegung plane, ${\bi v}\cdot{\bi r}^*_0={\bi v}\cdot{\bi u}^*_0=0$, and the CM is at the center of the circle and becomes equidistant of the trajectory of the CC. If ${\bi v}$, has a different orientation, because $R_0=1/2$, the separation between both centers CC and CM, $|{\bi r}-{\bi q}|\in[1/2-(v/2)\sin\alpha,1/2+(v/2)\sin\alpha]$, in natural units, is not a constant of the motion. 

We must remark that the CC ${\bi r}_0$ is the Poincar\'e transformed of the point ${\bi r}^*_0$, but 
${\bi q}_0$ is not the Poincar\'e transformed of the CM ${\bi q}^*_0$ in $O^*$. It is the CM of the particle in the laboratory frame $O$, at the same time $t_0$ than ${\bi r}_0$, while ${\bi r}^*_0$ and ${\bi q}^*_0$ are considered simultaneous in the frame $O^*$. Simultaneous events in one frame are not simultaneous in another. The point ${\bi q}_0$ represents the location of the CM of the particle simultaneous with ${\bi r}_0$, at the laboratory observer.
\section{Dirac particle in a uniform electric field}
\label{sec:uniformE}

The dynamical equation (\ref{eq:d2qdt2}) in natural units and in the presence of a uniform electric field along $OX$ axis ${\bi E}=E{\bi e}_x$ is
\[
\frac{d^2{\bi q}}{dt^2}=\frac{AE}{\gamma(v)}\left[{\bi e}_x-{\bi v}v_x\right],\quad A=\frac{e\hbar}{m^2c^3}=1\,{\rm n.u.}.
\]
If $E$ is expressed in V/m, we have to use the conversion factor (\ref{voltM}) and thus the parameter $AE$ is dimensionless. With a field of intensity 1 V/m the factor $a=AE=A\cdot1=7.55676\cdot10^{-19}$ in natural units.

In natural units, the system of differential equations is transformed into the system of first order
\[
\frac{d{\bi r}}{dt}={\bi u},\quad \frac{d{\bi u}}{dt}=\frac{1-{\bi v}\cdot{\bi u}}{({\bi q}-{\bi r})^2}({\bi q}-{\bi r}),
\]
\[
 \frac{d{\bi q}}{dt}={\bi v},\quad \frac{d{\bi v}}{dt}=\frac{aE}{\gamma(v)}\left[{\bi e}_x-{\bi v}v_x\right].
\]
This last equation written by components
\[
\frac{dv_x}{dt}=\frac{aE}{\gamma(v)}\left[1-v_x^2\right],
\]
\[
\frac{dv_y}{dt}=-\frac{aE}{\gamma(v)}\left[v_xv_y\right],
\]
\[
\frac{dv_z}{dt}=-\frac{aE}{\gamma(v)}\left[v_xv_z\right],
\]
where the value of $E$ is the intensity of the electric field in V/m.

\section{Dirac particle in an oscillating electric field}
\label{sec:oscE}
If in the above analysis we replace $E$ by a function $E(t)=E\cos(\omega t+\phi)$ we analyze the interaction of the Dirac particle with an oscillating electric field of angular frequency $\omega$ and initial phase $\phi$. The time has to be replaced by the dimensionless time so that the above field is
\[
E(t)=E\cos((\omega/\omega_0)t+\phi),\quad \omega_0=\frac{2mc^2}{\hbar}=1.55271\cdot 10^{21}{\rm s}^{-1}.
\]
With $\omega=\phi=0$, we obtain the previous example. The Mathematica notebook \cite{electric} produces the integration of both examples and the three components of the electric field can be selected. To obtain how the electric field affects to the CC and CM motion when we describe the particle in natural units we have to use very high electric fields and a small value for the dimensionless parameter $\omega/\omega_0$ in units of $\omega_0$, because the integration time corresponds to a value of $\pi$ during a turn of the CC around the CM. If the oscillating external electric field is of a GigaHertz frequency $10^9{\rm s}^{-1}$ the parameter $\omega/\omega_0\simeq10^{-12}$. For this value, the electric field during few turns of the CC, can be considered as a constant electric field and no visual effect can be appreciated. To appreciate the influence of the electric field at this scale we have to take a value of the parameter $\omega/\omega_0$ greater than this value.

\begin{figure}[!hbtp]\centering%
\includegraphics[width=7cm]{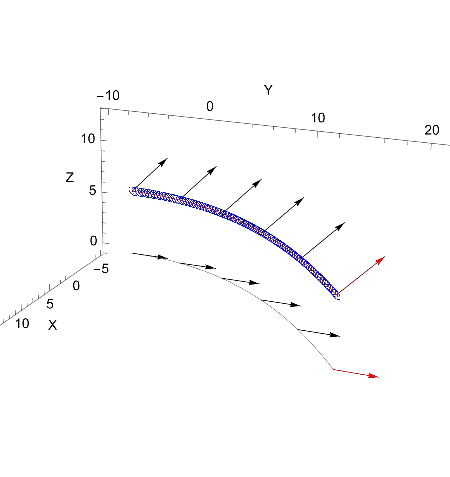}
\caption{Motion of the Dirac particle in a uniform electric field along the negative direction of the OX axis. We see the zitter motion of the CC (blue) and the motion of the CM (red). It is also depicted the CM spin, at different CM points, which forms an angle of $45^\circ$ with the direction of the initial CM velocity, of value $v_y=0.1$ in natural units. The interaction parameter $aE=1.133\cdot 10^{-3}$. The red spin is the last value of the spin at the end of the integration time. We also show the projection of the CM trajectory on the XOY plane and the projection of the CM spin, which at this low velocity and according to (\ref{variation}) remains almost constant. The spin of value 1/2 in natural units has been reescaled in the picture.} 
\label{CampoEx}
\end{figure}

In the figure {\bf\ref{CampoEx}} we represent the evolution of the Dirac particle in a uniform electric field ${\bi E}=-E{\bi e}_x$ of very high intensity to see in the same picture the zitter motion and how the external field bends the trajectory of the CM. The intensity of the electric field of this numerical integration is $E=3\cdot10^{15}$V/m and  the parameter $aE=1.133\cdot 10^{-3}$.

\begin{figure}[!hbtp]\centering%
\includegraphics[width=7cm]{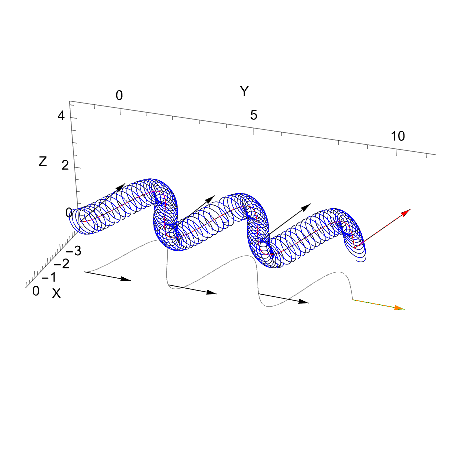}
\caption{Motion of the Dirac particle in an oscilating electric field along the direction of the OX axis of the same intensity as in the previous picture. We see the zitter motion of the CC (blue) and the motion of the CM (red). It is also depicted the CM spin, at different CM points, which forms an angle of $45^\circ$ with the direction of the initial CM velocity, of value $v_y=0.03$ in natural units. The natural frequency of the external field $\omega=0.056$. The integration time corresponds to a value of around $310$ units, approximately to 99 turns of the zitter motion. We also show the projection of the CM trajectory on the XOY plane and the projection of the rescaled CM spin.} 
\label{CampoExOscil}
\end{figure}

In the figure {\bf\ref{CampoExOscil}} we represent the integration in an oscillating electric field of the same intensity and with a value of $\omega/\omega_0=0.015$.

\section{Dirac particle in a uniform and oscillating magnetic field}
\label{sec:uniformB}
We are going to describe the motion of the Dirac particle, with the mass and charge of an electron, in a uniform magnetic field. To draw
in the same figure the cyclotron motion of the center of mass, that has a radius of order of $10^{-6}$m in a magnetic field of 5 Teslas, and the zitterbewegung motion of the CC around the CM of much smaller radius $R_0\simeq10^{-13}$m, we have to consider a high velocity electron and a very large magnetic field, to appreciate both trajectories in this figure.

Equation (\ref{eq:d2qdt2}) with ${\bi E}=0$, in natural units becomes:
\[
\frac{d^2{\bi q}}{dt^2}=\frac{A}{\gamma(v)}\left[{\bi u}\times{\bi B}c-{\bi v}\left(({\bi u}\times{\bi B}c)\cdot{\bi v}\right)\right],\quad A=\frac{e\hbar}{m^2c^3},
\]
with ${\bi B}$ in the international system of units. If the uniform magnetic field is along $OZ$ axis, ${\bi B}=B{\bi e}_z$, these equations become
\[
\frac{d^2{\bi q}}{dt^2}=\frac{KB}{\gamma(v)}\left[{\bi u}\times{\bi e}_z-{\bi v}({\bi u}\times{\bi e}_z)\cdot{\bi v}\right],\quad K=Ac=\frac{e\hbar}{m^2c^2}=2.2654\cdot10^{-10} {\rm T}^{-1}.
\]
 The system of dynamical equations is transformed into a system of first order differential equations:
\[
\frac{d{\bi r}}{dt}={\bi u},\quad \frac{d{\bi u}}{dt}=\frac{1-{\bi v}\cdot{\bi u}}{({\bi q}-{\bi r})^2}({\bi q}-{\bi r}),
\]
\[
 \frac{d{\bi q}}{dt}={\bi v},\quad \frac{d{\bi v}}{dt}=\frac{KB}{\gamma(v)}\left[{\bi u}\times{\bi e}_z-{\bi v}({\bi u}\times{\bi e}_z)\cdot{\bi v}\right].
\]
This last equation is explicitely written as:
\[
\frac{dv_x}{dt}=\frac{KB}{\gamma(v)}\left[u_y-v_x(v_xu_y-v_yu_x)\right],
\]
\[
\frac{dv_y}{dt}=\frac{KB}{\gamma(v)}\left[-u_x-v_y(v_xu_y-v_yu_x)\right],
\]
\[
\frac{dv_z}{dt}=\frac{KB}{\gamma(v)}\left[-v_z(v_xu_y-v_yu_x)\right],
\]
in natural units and the magnetic field is given in Teslas, so that the parameter $KB$ is dimensionless.

\begin{figure}
\hspace{2cm}\includegraphics[scale=0.5]{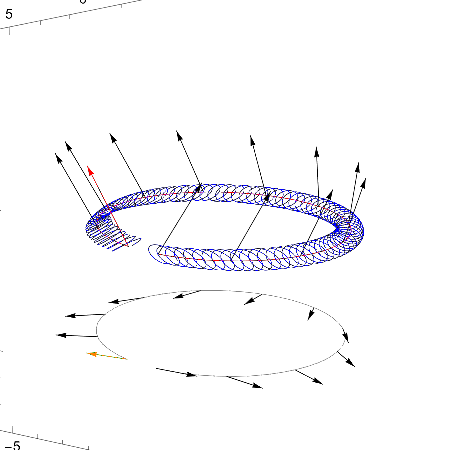}\includegraphics[scale=0.5]{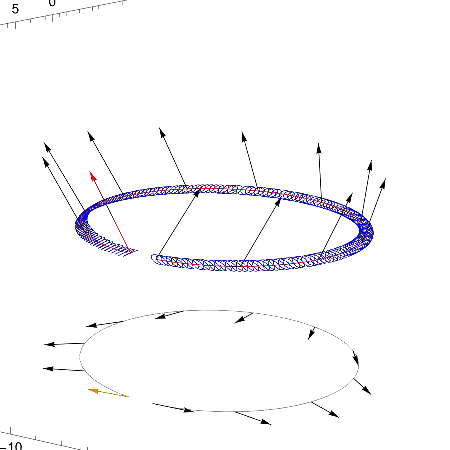}\includegraphics[scale=0.5]{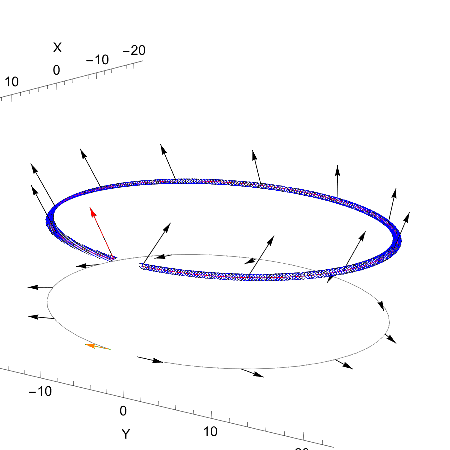}

\hspace{2cm}\includegraphics[scale=0.5]{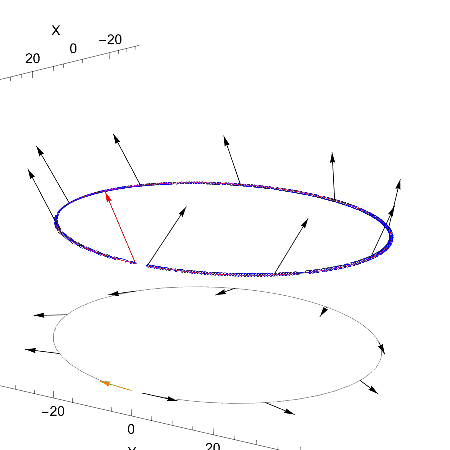}\includegraphics[scale=0.55]{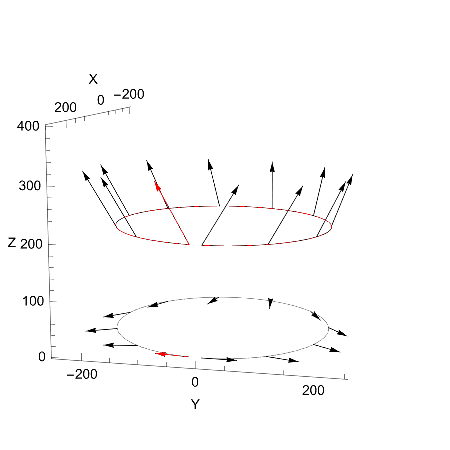}\includegraphics[scale=0.55]{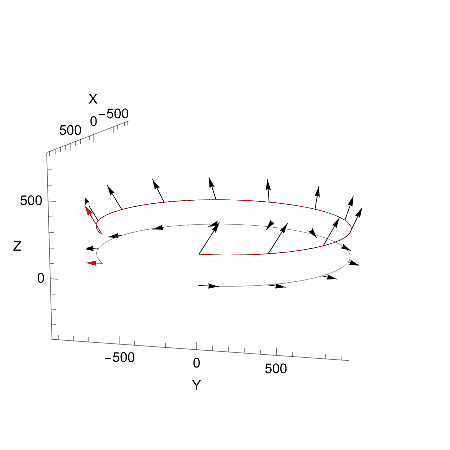}

\hspace{2cm}\includegraphics[scale=0.5]{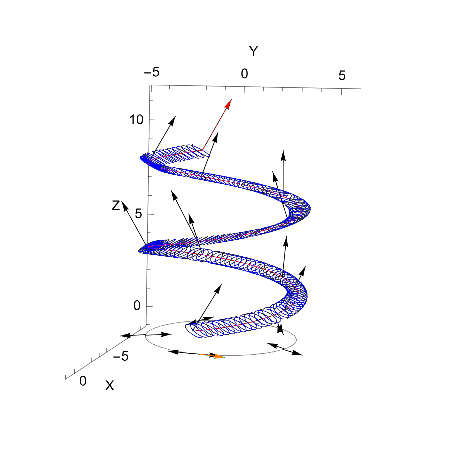}\includegraphics[scale=0.5]{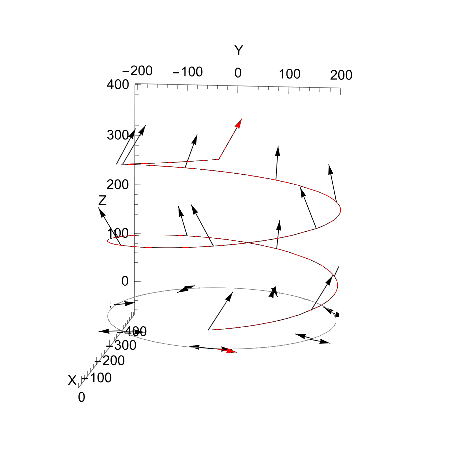}
\includegraphics[scale=0.7]{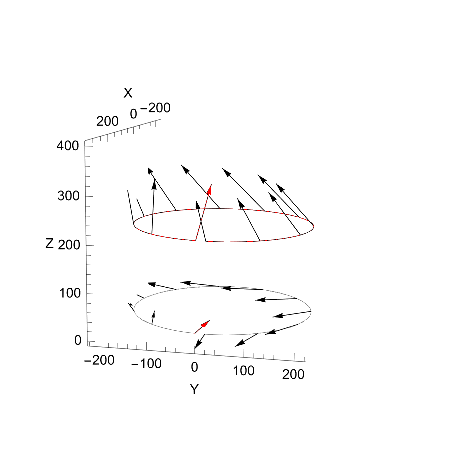}
\caption{Different cyclotron motions of radii $R_c=5,10,20,40,200$ and 800, with different values of the magnetic field and of the initial velocity of the CM. In all cases, except the last figure, the initial spin orientation is the same $\theta=30^\circ$, $\phi=90^\circ$, and always precesses backwards with Larmor's angular velocity. In the first four figures the motion of the CC is also visible at this scale but for larger cyclotron radii the motion of the CC is indistinguishable from the CM motion.\\ The first two figures of the third row show the two-turn cyclotron motion for two different radii with an initial component of the CM velocity along the magnetic field. After these two turns the CM spin returns to its initial orientation. The last figure in the third row is the same integration as the central figure but with a different orientation of the initial spin. After almost one turn the spin projection is opposite to the initial spin projection which justifies that the spin precesses backwards with $\omega_c/2$. The interested reader can modify the fields, the boundary conditions, the spin orientations and rotate the 3D figures to better appreciate the zitterbewegung and the cyclotron motion. The zitter motion of the CM is not noticeable at this scale, and the spin $S=1/2$ has been rescaled in every picture.}  
\label{fig:Rc5}
\end{figure}

In the Mathematica notebook \cite{electric} we analyze different examples of the Dirac particle in a uniform magnetic field, with different values of the external magnetic field, initial velocity of the CM of the particle and different spin orientations. 
Some of the bookmarks of the notebook include the integration of the examples depicted in figure {\bf\ref{fig:Rc5}}.

\begin{figure}[!hbtp]\centering%
\includegraphics[width=6cm]{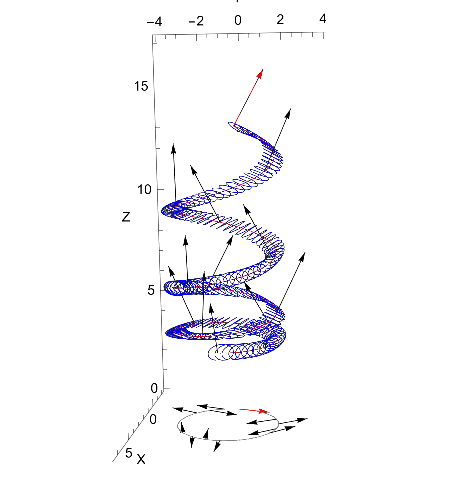}
\caption{Motion of the Dirac particle in a uniform magnetic field along OZ axis and a uniform electric field along the negative direction of the OZ axis, $B_z=2\cdot10^8$T, $E_z=-10^{14}$V/m. We see the zitter motion of the CC (blue) and the motion of the CM (red). It is also depicted the CM spin, at different CM points,with initial orientation of $\theta=30^\circ$ and $\phi=0$, with initial CM velocity, of value $v_x=0.12$ in natural units. We also show the projection of the CM trajectory on the XOY plane and the projection of the CM spin. The CM follows a cyclotron motion and is accelerated upwards. } 
\label{CampoEzB}
\end{figure}

In the figure {\bf\ref{CampoEzB}} we integrate the Dirac particle in a uniform magnetic field along the OZ axis and in a uniform electric field along the negative direction of the OZ axis of intensities  $B_z=2\cdot10^8$T, and $E_z=10^{14}$V/m, respectively. The CM has a cyclotron motion, and the zitter modification is negligible. The electric field produces an acceleration to the positive direction of OZ axis. The CM spin precess backwards with Larmor angular velocity.

\begin{figure}[!hbtp]\centering%
\includegraphics[width=6cm]{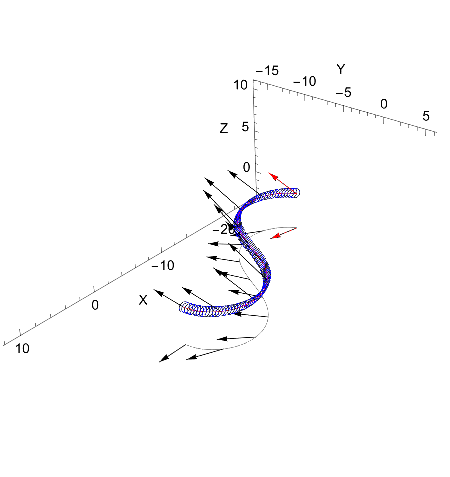}\includegraphics[width=6cm]{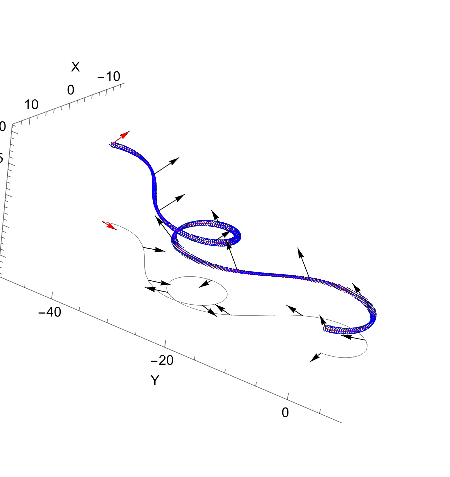}
\caption{Motion of the Dirac particle in an oscillating magnetic field along OZ axis  $B_z=10^8$T, and $\omega/\omega_0=0.0085$. We see the zitter motion of the CC (blue) and the motion of the CM (red). It is also depicted the CM spin, at different CM points, oriented an angle of $\theta=45^\circ$, $\phi=0$ with the initial CM velocity, of value $v_x=0.1$ in natural units. We also show the projection of the CM trajectory on the XOY plane and the projection of the CM spin. The right picture is the numerical experiment with the same intensity of $B$, $\omega/\omega_0=0.0056$, $v=0.11$ but forming an angle of $85^\circ$ with the OZ axis.} 
\label{oscilBz}
\end{figure}

In the figure {\bf\ref{oscilBz}} we analyze the motion of the Dirac particle in an oscillating magnetic field along OZ axis of value $B_z=10^8$T,  and natural frequency $\omega/\omega_0=0.009$. It is the same equation as above but if we replace in the differential equation the parameter $B$ by $B\cos(\omega t/\omega_0)$. It is also depicted the CM spin, at different CM points, which forms an angle of $45^\circ$ with the direction of the initial CM velocity, of value $v_x=0.12$ in natural units. The figure on the right is the same computation with the same oscillating magnetic field of the same internsity, frequency $\omega/\omega_0=0.0056$, $v=0.11$ but through an angle of $85^\circ$ with respect to the OZ axis. We see that the evolution of the spin is oscillating in a different direction when the curvature of the CM motion changes.\\ 

The mathematica notebook \cite{electric} has on the left panel the control of the boundary variables and also the possibility of auto-execute the computation by activating the corresponding button. In this case instead of depicting the whole solution, the program solves the dynamical equations in  steps by dividing the integration time in short periods. When the program reaches the final integration time, the program stops and integrates the equations again. This produces a film-like picture of the evolution of the Dirac particle. All the figures of this work appear as bookmarks of the mentioned notebooks.

\begin{figure}[!hbtp]\centering%
\includegraphics[width=7cm]{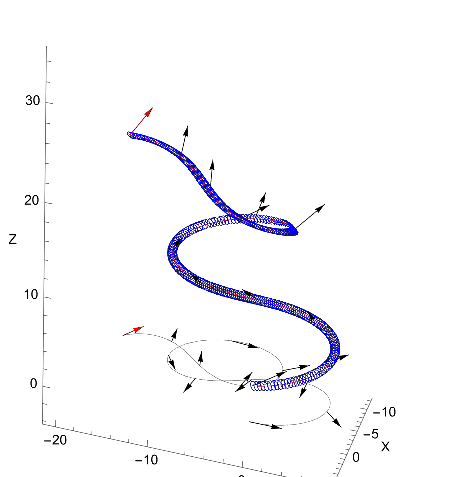}
\caption{Motion of the Dirac particle in an oscillating magnetic field along OZ axis  $B_z=9\cdot10^7$T, an oscillating electric field along OX, $E_x=6\cdot10^{14}$V/m and $\omega/\omega_0=0.0056$. We see the zitter motion of the CC (blue) and the motion of the CM (red). It is also depicted the scaled CM spin, at different CM points, which forms an angle of $25^\circ$ with the direction of the initial CM velocity, of value $v=0.11$ in natural units oriented at $85^\circ$ with respect to the OZ axis. We also show the projection of the CM trajectory on the XOY plane and the projection of the CM spin.} 
\label{oscilBzEx}
\end{figure}

In the figure {\bf\ref{oscilBzEx}} we describe the evolution of the Dirac particle in an oscillating electric and magnetic field $E_x$ and $B_z$ with an initial velocity forming an angle of $5^\circ$ with the OY axis.
\newpage

\section{Conclusions}
The analytical description of the classical Dirac particle is determined by the evolution of a single point ${\bi r}$, interpreted as the center of charge, that satisfies a system of ordinary fourth-order differential equations, in such a way that all observables can be defined in terms of this point and the subsequent time derivatives.
This particle has two distinguished points, the center of mass ${\bi q}$ and the center of charge ${\bi r}$, where the CC velocity ${\bi u}$ is $u=c$ and the CM velocity ${\bi v}$ satisfies $v<c$. The motion of the CC is, in general, a curve with torsion what is known in the literature as the {\it zitterbewegung} motion. These fourth-order differential equations for point ${\bi r}$, can be decoupled into a system of second-order differential equations for the points ${\bi r}$ and ${\bi q}$.

Because we have two distinguished points we have two different spin observables with respect to both points. The CC spin ${\bi S}$, is the classical equivalent to the quantum mechanical Dirac spin operator, and it is not conserved even in the free motion. It evolves in a kind of precession around the CM spin ${\bi S}_{CM}$, such that its time derivative $d{\bi S}/dt={\bi p}\times{\bi u}$, satisfies the same equation as Dirac's spin operator in the quantum case, it is orthogonal to the CC velocity ${\bi u}$ and the linear momentum ${\bi p}$. In this way the CM spin, that is conserved in the free motion, is like some average of the CC spin during one zitterbewegung turn of the motion of the CC. This internal motion has an internal frequency in the CM frame $\nu_0=2mc^2/h$, that defines a natural unit of time. When the CM is moving the internal frequency decreases as $\nu(v)=\nu_0/\gamma(v)$.

The linear momentum and energy are expressed in terms of the CM velocity like in the point particle case ${\bi p}=\gamma(v)m{\bi v}$ and $H=\gamma(v)mc^2$, respectively. The Hamiltonian is the classical equivalent of Dirac's Hamiltonian and is interpreted as the sum of the translation energy of the CM and the rotation energy, which never vanishes.

The particle has an arbitrary mass $m$ and arbitrary electric charge $e$. The classical spin can take any arbitrary value $S$, but when quantized the model the classical parameter $S=\hbar/2$.

In Section {\bf\ref{interaction}} we have analyzed the local evolution of the spins and angular velocity observables in the free case, at this natural time scale. Two different motions have been described whether the CC trajectory is flat or it has torsion. If the CC trajectory is flat the angular velocity has a vanishing component $\bomega_u$ along the velocity of the CC ${\bi u}$. Even when the CC trajectory is flat and the motion is free, the ${\bi S}_{CM}$ is conserved but the CC spin ${\bi S}$ oscillates around ${\bi S}_{CM}$. This spin ${\bi S}$ is only conserved in the free case and for the center of mass observer where ${\bi p}=0$.

The invariant properties of the Dirac particle are the absolute value of the two four-vectors $p^\mu\equiv (H/c,{\bi p})$ and the Pauli-Lubanski four-vector $w^\mu\equiv({\bi p}\cdot{\bi S}_{CM},H{\bi S}_{CM})$ expressed in terms of the energy $H$, linear momentum ${\bi p}$ and the CM spin ${\bi S}_{CM}$. These define $p^\mu p_\mu=m^2c^2$ and $w^\mu w_\mu=-m^2c^2S(0)^2$, where $S(0)=\hbar/2$ is the absolute value of the spin in the center of mass frame. The intrinsic properties are $m$ and $S(0)$. From the transformation properties of the Pauli-Lubanski four-vector we have obtained how the spin transforms among inertial observers. The size of the spin in an inertial reference frame depends on the relative velocity between this frame and the center of mass frame of the Dirac particle and decreases at very high velocity.

The whole formalism can be described in dimensionless variables and natural units. The proposed natural system of units takes as fundamental units: the unit of length is twice the separation between the CC and the CM, $2R_0$. The unit of velocity is $c=1$. The unit of time is $\tau_0=2R_0/c=1$. The unit of action is Planck's constant $\hbar=1$ which defines that the unit mass of the electron is $m=1$. Finally the unit of electric charge is $e=1$ which means that the permitivity of the vacuum $1/4\pi\epsilon_0=\alpha$, where $\alpha$ is the fine structure constant.

For the electron and positron, the intrinsic magnitudes are in natural units: mass $m=1$, the separation between CC and CM $R_0=1/2$, spin $S=mR_0c=1/2$, electric charge $e=\pm1$, electric dipole moment $d=eR_0=1/2$, and magnetic dipole moment $\mu_B={e\hbar}/{2m}=1/2$. The expressions in natural units of the energy $H=\gamma(v)$, linear momentum ${\bi p}=\gamma(v){\bi v}$,  CC spin ${\bi S}=-\gamma(v)({\bi r}-{\bi q})\times{\bi u}$, CM spin ${\bi S}_{CM}=-\gamma(v)({\bi r}-{\bi q})\times({\bi u}-{\bi v})$ and magnetic moment
$\bmu_{CM}=({\bi r}-{\bi q})\times({\bi u}-{\bi v})$, with the opposite sign for the positron.

We have explored the interaction of the spinning Dirac particle with uniform and oscillating electric and magnetic fields. We use the minimal coupling interaction and the classical parameters $m$ and $e$, are those of the electron.
The analysis is a kind of microscopic analysis at the scale of the separation between the CC and the CM.

The motion of the center of mass of the particle under these fields, seems to be the expected motion similar to the motion of the point particle with a slight zitter. The new feature has been to determine how the spin evolution is affected by the interaction. To depict in the same figure the global and the internal motion we have been concerned with the use of very high external fields. The numerical analysis has been performed with very high hypothetical fields that are difficult to produce at the laboratory. Since the dynamical equations are relativistically invariant they can be solved for arbitrary high velocities. If we use laboratory fields we will not distinguish at that scale a difference between the CC and CM motions, like the graphics of figure {\bf\ref{fig:Rc5}}, but we can always analyze the spin evolution.

The reason is the existence of two different scales: the scale of the separation beteween the CM an CC, of order of $R_0\approx10^{-13}$m, to show the relative motion between these two points, and the macroscopic scale at the laboratory level.
The other is the time scale. The CC is turning around the CM at the speed of light and in natural units the time taken during a turn is $\pi\gamma(v)$ where $v$ is the velocity of the CM ${\bi q}$, and this period is of order of $10^{-21}$s, for low velocity Dirac particles and multiplied by the $\gamma(v)$ factor for high energy particles. 

The uniform and oscillating electric field seems not to affect the evolution of the CM spin ${\bi S}_{CM}$ that remains almost constant during a few periods of the relative internal motion at the analyzed low velocity. The magnetic field produces the precession of the ${\bi S}_{CM}$ with the Larmor angular velocity, and this spin precesses backwards, in the opposite direction to the angular velocity of the CM motion, such that when the CM has given two turns the spin comes back to its initial orientation. In the interaction with an oscillating magnetic field, each time the magnetic field reverses, the angular velocity of the CM reverses and also the Larmor angular velocity of the center of mass spin. Combinations of oscillating electric and magnetic fields produce the modification of the CM spin orientation, although in these examples the CM velocity is not high and the absolute value of the spin is not modified.

The analysis of the interaction of the Dirac particle with an electromagnetic plane wave is left to a future paper.

We have not explored the possible interaction of the Dirac particle with external fields where the scalar and vector functions $A_0$ and ${\bi A}$, respectively, could also be functions of the CC velocity ${\bi u}$. This analysis would produce a classical non-electromagnetic interaction which to our knowledge has never been analyzed and deserves a further research.

The notebook \cite{electric} can be executed in auto mode to obtain a film-like picture of the numerical integration in uniform or oscillating electric and magnetic fields. For other kind of fields you can modify the fields in the dynamical equations of the numerical code of the notebook at your will. The analytical expressions of the fields have to be written in natural units. This notebook contains as bookmarks all figures of this article. If the auto mode is suppressed all boundary conditions can be modified manually to obtain the different interaction scenarios of the classical Dirac particle in an external electromagnetic field.


\end{document}